\documentclass[12pt,onecolumn]{IEEEtran}

\usepackage{amsmath}
\usepackage{amsfonts}
\usepackage{theorem}
\usepackage[dvips]{epsfig}      
\usepackage{graphicx}
\usepackage{enumerate}
\usepackage{caption}
\usepackage{subcaption}
\usepackage{thmtools}
\usepackage{amsfonts} 
\usepackage{verbatim}
\usepackage{amssymb}
\usepackage{amsbsy}
\usepackage{setspace}
\usepackage{url}
\usepackage{lineno}
\usepackage{multicol}

\theoremstyle{plain}
\newtheorem{Theorem}{Theorem}
\newtheorem{Lemma}{Lemma}
\newtheorem{Proposition}{Proposition}
\newtheorem{Corollary}{Corollary}

\newtheorem{Problem}{Problem}

{\theorembodyfont{\rmfamily} \newtheorem{Remark}{Remark}}
{\theorembodyfont{\rmfamily} }
{\theorembodyfont{\rmfamily} \newtheorem{Example}{Example}}

{\theorembodyfont{\rmfamily} }
{\theorembodyfont{\rmfamily} }

\newcommand {\R}{\mathbb R}

 \newcommand{\Int}{\operatorname{int}}

\newcommand{\be}{\begin{equation}}
\newcommand{\ee}{\end{equation}}

\newcommand{\diag}{\operatorname{{\mathrm diag}}}

\newcommand{\rfmfull}{\textsl{RFM with different site sizes}}
\newcommand{\rfmrfull}{\textsl{RFMR with different site sizes}}

\newcommand{\rfmshort}{\textsl{RFMD}}
\newcommand{\rfmrshort}{\textsl{RFMRD}}
\newcommand{\prfmrshort}{\textsl{PRFMRD}}

\newcommand{\LMDR}{\lambda_1,\dots,\lambda_n}

\newcommand{\LMDRR}{\lambda_1,\dots,\lambda_{n+2}}
\newcommand{\QR}{q_1,\dots,q_n}

\newcommand{\QRR}{q_1,\dots,q_{n+2}}

\newcommand*\diff{\mathop{}\!\mathrm{d}}

\begin{document}
%

\title{Ribosome flow model with different site sizes }

\author{Eyal Bar-Shalom and Alexander Ovseevich and Michael Margaliot 
 \IEEEcompsocitemizethanks{
\IEEEcompsocthanksitem
  E. Bar-Shalom is with the School of Electrical Engineering, Tel-Aviv University, Tel-Aviv 69978, Israel.
E-mail:  eyalbarshalom500@gmail.com
\IEEEcompsocthanksitem A. Ovseevich is with the Ishlinsky Institute for Problems in Mechanics,
Russian Academy of Sciences and the Russian Quantum Center, Moscow, Russia. E-mail: ovseev@ipmnet.ru
\IEEEcompsocthanksitem
M. Margaliot  (corresponding author) is with the School of Electrical Engineering and the Sagol School of Neuroscience,
Tel-Aviv University, Tel-Aviv 69978, Israel.
E-mail: michaelm@eng.tau.ac.il  
}}


\maketitle
 \doublespace 

\begin{abstract}

We introduce and analyze two general
dynamical models for unidirectional movement of particles along 
a circular chain and an open chain of sites. The models include a soft version of the 
 simple exclusion principle, that is,  as the density in a site increases the effective entry rate into this site decreases. This allows to model and study the evolution of ``traffic jams'' of particles
along the chain.
A unique feature of these two new models is that each site along the chain can have a different size.

Although the models are nonlinear, they are   amenable to rigorous asymptotic
 analysis. In particular, we 
  show that the  
dynamics always converges to a steady-state, and that the steady-state densities along the chain 
and the steady-state  output  flow rate  from the chain 
can be derived from the spectral properties of a suitable matrix, thus eliminating the need to numerically simulate the dynamics until convergence. This spectral representation also allows for powerful sensitivity analysis, i.e. understanding how a change in one of the parameters in the models affects the steady-state. 

We show  that the site sizes and the transition rates from site to site
 play different roles in the dynamics, and that 
for the purpose of maximizing the steady-state 
output (or production) rate the site sizes are more important than the transition rates. We also show that the problem of finding parameter values that maximize
 the production rate is tractable.

We believe that the models introduced here can be applied to study various natural and artificial processes including ribosome flow during mRNA translation, 
the movement of molecular motors along filaments of the cytoskeleton, 
pedestrian and 
vehicular traffic, evacuation dynamics, and more. 
\end{abstract}
 

\section{Introduction} \label{sec:int} 

Understanding various transport phenomena 
in the cell is of considerable interest.
Fundamental cellular processes like
transcription, translation, and the 
movement
 of molecular motors can be 
studied  using a general  
model for the flow of ``particles'' along 
a cellular ``track''. The particles may be ribosomes moving along the mRNA strand  or
molecular motors moving along  actin filaments.
To   increase the flow, often several particles 
traverse     the same   track simultaneously. 
For example, during mRNA translation several ribosomes may ``read''
 the same mRNA strand simultaneously (thus forming a polysome). It is
important to note that
new experimental methods
are providing unprecedented 
  data  on the
dynamics of this 
fundamental biological process~\cite{CHEKULAEVA2016918},
thus increasing the interest in computational models that can integrate and explain this data.

A simple physical concept
 underlying such motion is the \emph{simple exclusion principle}: two particles cannot be in the same site along the track  at the same time. This implies that a ``traffic jam''
of particles 
may evolve behind a particle
that remains in the same site  
 for a long time. The evolution and  implications of such traffic jams
in various biological processes are  attracting  considerable interest (see, e.g.~\cite{Ross5911,tuller_traffic_jams2018,neurojams}).

To study the  transport phenomena
 in the cell in a qualitative and quantitative manner, scientists build computational models,
identify  useful control parameters,  and determine the functional dependence of the transport properties on these parameters.
Such models are particularly important in the context
of synthetic biology and    biomimetic systems where   biological 
modules are modified or redesigned~\cite{DELVECCHIO20185}.
An important goal in such studies is to
 determine how 
the density of particles along the chain 
depends on the structure
 and parameters of the system, 
and to find  parameter values that 
lead to an optimal production
rate~\cite{RFM_r_max_density,KLUMPP20053118,Leduc6100,rfm_down_regul}.

A fundamental model from statistical physics is the
 \emph{totally asymmetric simple exclusion process}~(TASEP)~\cite{Shaw2003,TASEP_tutorial_2011,MALLICK201517}. 
This  is a   stochastic model for unidirectional movement that takes place on some kind of tracks or trails. The tracks are modeled by an ordered
 lattice of sites, and the moving objects are modeled as
  particles that can hop, with some probability, from one site to the consecutive site. The motion is assumed to be asymmetric in the sense that there is some preferred direction of motion. The term \textsl{totally asymmetric} refers to the case where motion is unidirectional. The term \emph{simple exclusion} refers to the fact that hops to a target site may take place only if it is not already occupied by another particle. 
Note that every site may either by empty
or contain a single particle, so in particular all the
sites have the same  size.

TASEP has two basic configurations, open boundary conditions and periodic boundary conditions. In the first configuration, the lattice boundaries are open and the first and last sites are connected to external particle reservoirs. In TASEP with periodic boundary conditions, the lattice is closed, so that a particle that hops from the last site returns back to the first one. Thus, the particles hop around a circular chain, and the total
number of particles along  the lattice is conserved.

In this paper, 
we introduce and rigorously analyze two nonlinear continuous-time  dynamical models describing
the unidirectional movement of ``particles'' along 
a circular and an open chain of~$n$ sites. 
For every index~$i\in\{1,\dots,n\}$ site~$i$ has a size site (i.e. maximal possible capacity)~$q_i$,
and the transition to site~$i+1$ is controlled by a parameter~$\lambda_i$. The state-variable~$x_i(t)$, that takes values in~$[0,q_i]$, describes the density of particles at site~$i$ at time~$t$.
The models include a soft version of
the \emph{simple exclusion principle}. 
This allows to study the evolution of ``traffic jams''
along the chain and, in particular, the effect of  a small transition rate~$\lambda_i$ or a small site size~$q_i$. 
A unique feature of these models is that each site along the chain can have a different size.
Indeed, there is no a priori reason to expect that the capacity   in  two different sites is equal. For example, if we consider the flow of vehicular 
 traffic along a road then the capacity     changes when the number of parallel lanes along the road  increases or decreases.

Although nonlinear, the new models are amenable to rigorous analysis. 
Our   results
 show that the dynamics always converges to a steady-state.
In other words, 
as time goes to infinity, the density~$x_i(t)$ at every site~$i\in\{1,\dots,n\}$ 
converges to a steady-state value~$e_i$, with~$e_i \in [0,q_i]$.
This means that as time goes to infinity, the effective entry rate into site~$i$
and the effective exit rate from site~$i$ become equal, yielding a constant density~$e_i$
at site~$i$. 
In the open  chain,
  these steady-state densities depend on all the parameters~$q_i,\lambda_i$, but not on the initial density~$x_j(0)$, $j=1,\dots,n$,
	at each site.
	In the circular model, the steady-state densities depend on all the parameters~$q_i,\lambda_i$, and also on the initial
	total    density, i.e.~$x_1(0)+\dots+x_n(0)$ along the chain.

Surprisingly, we show that in  both models the steady-state densities and flow rate can be derived from 
 the spectral properties of a suitable matrix, thus eliminating the need to numerically simulate the dynamics until convergence. This spectral representation also allows a
 powerful sensitivity analysis, i.e. understanding how a change in one of
the parameters in the models affects the steady-state. 
Furthermore, we apply  the
 spectral representation   to show that
the mapping from the model parameters to the steady-state flow rate
is quasi-concave implying that the problem of maximizing the flow rate
is numerically tractable  even for very long chains.

The remainder of this paper is organized as follows. The next section reviews several related models 
 and in particular emphasizes the unique features
of the  new  models introduced here.
Section~\ref{sec:tmodels} 
describes the two new models  
for movement along a circular and  an open chain.
The main analysis results are described in 
Sections~\ref{sec:main} and~\ref{sec:main_open}. 
We first analyze the circular model and then show that 
  the steady-state behavior
in the  open    model can be derived 
by taking one of the transition rates~$\lambda_i$ 
in the~$n$-dimensional 
circular model to infinity. This effectively ``opens the loop'' in the circular model   
 yielding  an open   model with dimension~$n-2$. 
The final section concludes and describes several
 directions for further research. 
 
\section{Preliminaries}\label{sec:pre}

The \emph{ribosome  flow model}~(RFM)~\cite{reuveni} is a dynamic mean-field approximation of~TASEP with open boundary conditions. The~RFM has been extensively used
to model and analyze 
 ribosome  flow along  an mRNA molecule~\cite{RFM_entrain,HRFM_steady_state,RFM_stability,RFM_feedback,rfm_max,zarai_infi,HRFM_concave,RFM_r_max_density}.
The molecule is    coarse-grained into~$n$  codons (or groups of codons). Ribosomes reach the first site with initiation rate~$\lambda_0>0$, but the effective entry rate
decreases as the density in the first site increases.
 A ribosome that occupies site~$i$ moves, with transition rate $\lambda_i>0$, to the consecutive site but again the effective rate decreases as the
consecutive site becomes more occupied.

The \emph{ribosome flow model on a ring}~(RFMR)~\cite{RFMR,alexander2017} is the dynamic mean-field 
of~TASEP with periodic boundary conditions. 
Here the  particles exiting the last site enter the first site.  The~RFMR dynamics
admits a first integral as  the total density along the chain is preserved. 
The~RFMR has been used as a model for mRNA translation with ribosome recycling. Note that
a recent study~\cite{Morisaki1425} 
  concluded that polysomes
are globular in shape rather than elongated, based on
the observation that the distance between protein- and mRNA-labeling fluorophores was largely unaffected by the length of the
coding sequence.

In both the RFM and RFMR all the sites along the chain are assumed to have the same size, 
 and this is normalized to one. 
Here, we introduce and analyze generalizations of these models, called the~{\rfmfull} and~{\rfmrfull}, respectively, that
allow for different site sizes.

\section{  New Models}\label{sec:tmodels} 
\begin{figure}
 \begin{center}
  \includegraphics[scale=0.8]{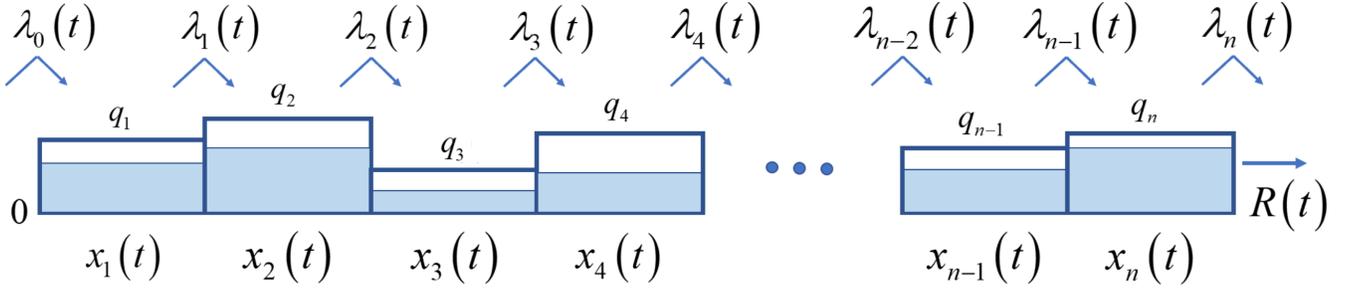}
	\caption{The {\rfmfull} models unidirectional flow along a chain of~$n$ sites. The state variable~$x_i(t)\in[0,q_i]$ represents the density at site $i$ at time $t$. The maximal possible density at site~$i$ is~$q_i$. The parameter $\lambda_i>0$ controls the transition rate from  site~$i$ to site~$i+1$, with~$\lambda_0>0$ [$\lambda_n>0$] controlling the initiation [exit] rate. The output rate at time $t$ is~$R(t) :=\lambda_n x_n(t)$. }
	\label{fig:rfm_diff_cell_size_model_fig}
\end{center}
\end{figure}

We begin with  the open model, i.e. the {\rfmfull}~({\rfmshort}) depicted in Fig.~\ref{fig:rfm_diff_cell_size_model_fig}.
This is described by~$n$ first-order differential equations:
\begin{align}\label{eq:rfmrd}
                    \dot{x}_1&=\lambda_0 (q_1-x_1) -\lambda_1 x_1(q_2-x_2), \nonumber \\
                    \dot{x}_2&=\lambda_{1} x_{1} (q_2-x_{2}) -\lambda_{2} x_{2} (q_3-x_3) , \nonumber \\
                             &\vdots \nonumber \\
                    \dot{x}_n&=\lambda_{n-1}x_{n-1} (q_n-x_n) -\lambda_n x_n,
\end{align} 
with~$\lambda_i>0$ and~$0<q_i \leq 1$ for all~$i$. 
The state variable $x_i(t): \R_+ \rightarrow [0,q_i]$, $i=1,...,n$, describes the normalized   occupancy level at
 site~$i$ at time~$t$, where $x_i(t)=q_i$  $[x_i(t)=0]$ indicates that site~$i$ is completely full [empty] at time $t$. 

The model includes $2n+1$ positive parameters. 
The  
parameters~$\lambda_0,\dots,\lambda_n$ describe the maximal possible transition rate between the sites: the initiation rate~$\lambda_0$ into the chain, 
the elongation (or transition) rate~$\lambda_i$ from site~$i$ to site~$i+1$, $i=1,...,n-1$, and the exit rate~$\lambda_n$.
The   parameters~$q_1,\dots,q_n\in(0,1]$  describe the maximal capacity at
 each site. The use of different values~$q_i$ allows to model flow through a chain of sites with different sizes. In the special
 case where $q_i = 1$  for all~$i=1,\dots,n$  we retrieve the~RFM that  has been
extensively  used to model and analyze the flow of ribosomes along the mRNA molecule during translation (see, e.g.~\cite{reuveni,Zarai20170128,RFM_model_compete_J,rfm_down_regul}). 

It is important to note that the {\rfmshort} cannot be derived by simply scaling the
state-variables in the~RFM. The next example demonstrates this.  
\begin{Example}\label{exa:rfm_scale}
Consider an RFM with~$n=2$, i.e.
\begin{align*}
\dot x_1&=\lambda_0(1-x_1)-\lambda_1 x_1(1-x_2),\\
\dot x_2&=\lambda_1 x_1(1-x_2)-\lambda_2 x_2.
\end{align*}
Define new state-variables~$z_i(t):=s_ix_i(t)$, with~$s_i>0$. Then the equations in the new state-variables are:
\begin{align}\label{eq:zeqns}
\dot z_1&=\lambda_0(s_1-z_1)-\frac{\lambda_1}{s_2} z_1(s_2-z_2),\nonumber \\
\dot z_2&=\frac{\lambda_1}{s_1} z_1(s_2-z_2)-\lambda_2 z_2.
\end{align}
If~$s_1 \not =s_2$ then~\eqref{eq:zeqns}
  is \emph{not} an {\rfmshort}, as the flow out of site~$1$   is
$\frac{\lambda_1}{s_2} z_1(s_2-z_2)$ whereas
 the flow into site~$2$ is~$\frac{\lambda_1}{s_1} z_1(s_2-z_2)$, and these are not equal. 
If~$s_1   =s_2$ then~\eqref{eq:zeqns} 
  is also \emph{not} a general~{\rfmshort}, as both sites have the same size, namely,~$s_1=s_2$. 
\end{Example}


The different site sizes 
  in the~{\rfmshort} add  important dynamical features that do not exist in  the~RFM nor other equal-site models
	like~TASEP.
The next example demonstrates this. 

\begin{figure} 
\centering
\begin{subfigure}{.55\textwidth}
  \centering
  \includegraphics[width=.8\linewidth]{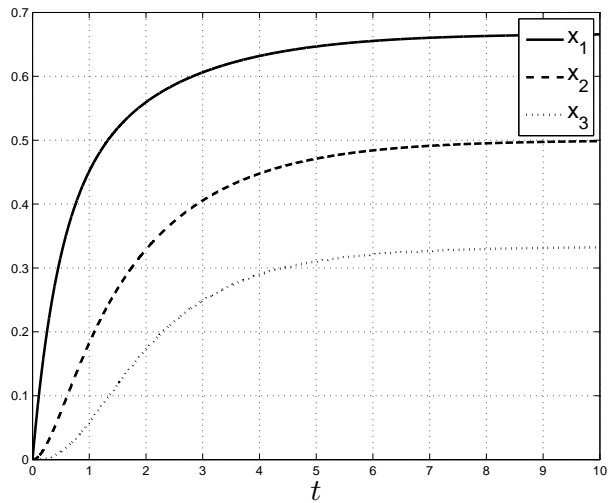}
  \caption{ RFM}
  \label{fig:sub1}
\end{subfigure}%
\begin{subfigure}{.55\textwidth}
  \centering
  \includegraphics[width=.8\linewidth]{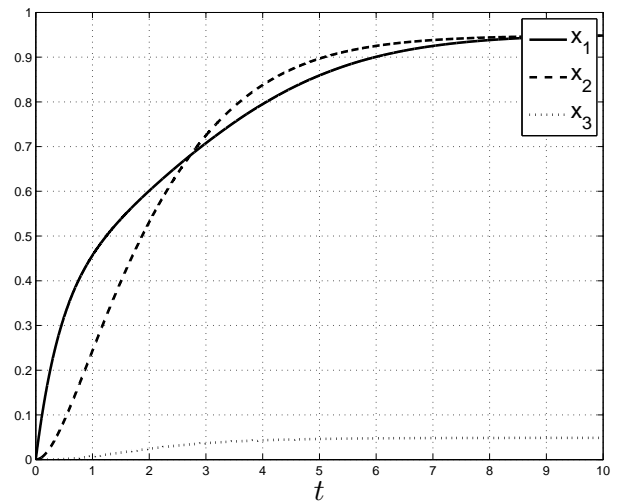}
  \caption{{\rfmshort}  with $q_1=q_2=1, q_3=0.1$}
  \label{fig:sub2}
\end{subfigure}
\caption{State-variables~$x_i(t)$ in an RFM (left) and an  {\rfmshort} (right), both of dimension~$n=3$, as a function of time. 
In both models~$\lambda_i=1$, $i=1,2,3$. 
}
\label{fig:test}
\end{figure}

\begin{Example}
Fig.~\ref{fig:test} depicts the state-variables~$x_i(t)$, $i=1,2,3$,
in an {\rfmshort} with~$n=3$
and compares them to  the state-variables 
in an~RFM with~$n=3$. 
In both models all the~$\lambda_i$'s are set to one. 
In the~{\rfmshort} the site sizes are~$q_1=q_2=1$, and~$q_3=0.1$.
Thus, the last site has a much smaller size than the first two. 

It may be seen that in both models the state-variables converge to a steady-state. However,  the steady-state behavior in the two models is quite different. 
The small size of site~$3$ in the~{\rfmshort} 
makes it fill up quickly.
Consequently, site~$2$ fills up and then also site~$1$. 
This generates a  ``traffic jam'' in the~{\rfmshort}. Thus,
in the~{\rfmshort} there can be  two different ``bottlenecks'' that generate traffic jams: a small transition rate or a small site size. 
\end{Example}

We now turn to describe the~{\rfmrshort}.
This is similar to the~{\rfmshort}, but 
under   the  additional assumption  that
all the particles leaving site~$n$ circulate back to site~$1$. 
The equations are thus:
\begin{align}\label{eq:rfmrl}
                    \dot{x}_1&=\lambda_n x_n (q_1-x_1) -\lambda_1 x_1(q_2-x_2), \nonumber \\
                    \dot{x}_2&=\lambda_{1} x_{1} (q_2-x_{2}) -\lambda_{2} x_{2} (q_3-x_3) , \nonumber \\
                             &\vdots \nonumber \\
                    \dot{x}_n&=\lambda_{n-1}x_{n-1} (q_n-x_n) -\lambda_n x_n (q_1-x_1) .
\end{align} 
Note that here the entry rate into site~$1$ is equal to the exit rate from site~$n$. This models a flow of particles along a circular chain, rather than an 
open chain. 
When considering the~{\rfmrshort} we always interpret the indexes modulo~$n$.
For example,~$\lambda_{n+1}=\lambda_1$ and~$q_0=q_n$.

\begin{figure}
 \begin{center}
  \includegraphics[scale=0.8]{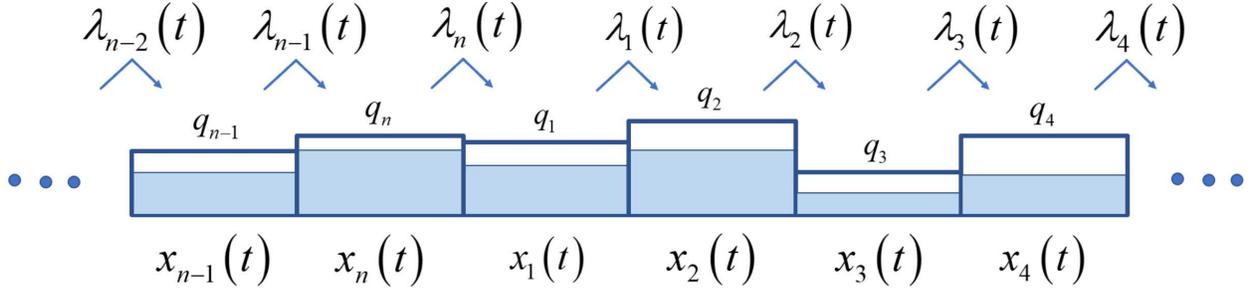}
	\caption{The {\rfmrshort} models unidirectional flow of particles along a circular chain of~$n$ sites. The maximal  possible density
	in site $i$ is $q_i\in(0,1]$. The parameter $\lambda_i>0$ controls the transition rate from site~$i$ to site~$i+1$.}
  \label{fig:rfmr_diff_cell_size_model_fig}
\end{center}
\end{figure}

In the special case where~$q_i=1$ for all~$i$ the~{\rfmrshort} in~\eqref{eq:rfmrl}
becomes the ribosome flow model on a ring~(RFMR) that has been used to study 
ribosome flow with circularization~\cite{RFMR,alexander2017,RFM_r_max_density}.

The next two
sections describe  the mathematical properties of the new models.
We begin by analyzing the~{\rfmrshort}, as we will later
show that the theoretical
 results for the~{\rfmshort} follow by taking~$\lambda_n\to \infty$ in an~{\rfmrshort} with a specific total   density. 
 To increase readability, all the proofs are placed
in the Appendix.

\section{Analysis of the {\rfmrshort}}\label{sec:main}
The state space of~\eqref{eq:rfmrl} is the set
$C:=[0,q_1]\times\dots\times [0,q_n]
$.
For any initial condition~$a \in C$, let~$x(t,a)$ denote the solution  at time~$t$ of~\eqref{eq:rfmrl} 
with~$x(0)=a$.
Define the function~$H:\R^n_+ \to \R_+$ by
$
H(y):=y_1+\dots+y_n.
$ 
An important property of~\eqref{eq:rfmrl} 
is that~$\sum_{i=1}^n \dot x_i(t) \equiv 0$.
 This means that  along any solution of~\eqref{eq:rfmrl} we have 
\[
H(x(t,a)) \equiv  H(a).
\] 
In other words, the total density along the circular chain is conserved. For~$s \in [0, q_1+\dots+q_n]$, let
\[
			L_s:=\{y\in C : \sum_{i=1}^n y_i = s \}
\]
denote the~$s$ level-set of~$H$, i.e. the set of all points~$y\in C$ such that~$H(y)=s$.
For example, for~$n=2$ and~$s=3$ 
the set~$L_3$ includes the points~$\begin{bmatrix} 0& 3\end{bmatrix}^T$,
  $\begin{bmatrix} 0.5& 2.5\end{bmatrix}^T$,   and so on.  

\subsection{Invariance and asymptotic stability}
The next result shows that for any~$a\in C$ the solution~$x(t,a)$ of the~{\rfmrshort}
remains in~$C$ for all~$t\geq 0$. In other words,
for any~$i$, the density~$x_i(t) \in[0,q_i]$
for any time~$t\geq 0 $.  This means that 
the density remains  well-defined for all~$t\geq 0 $. 
Furthermore,~$x(t,a)$ converges to a steady-state that depends on the~{\rfmrshort}
parameters and on the initial total density~$x_1(0)+\dots+x_n(0)$. 
Recall that all the proofs are placed in the Appendix. 

 \begin{Proposition} \label{prop:sta_inv}
The set~$C$ is an invariant set of~\eqref{eq:rfmrl}. 
For any~$s\in[0,q_1+\dots+q_n]$ the set~$L_s$ includes a unique
 steady-state~$e^s$
and any solution~$x(t,a)$
 of~\eqref{eq:rfmrl} with~$\sum_{i=1}^n a_i(0)=s$
satisfies
\[
					\lim_{t\to \infty} x(t,a)=e^s.
\]
\end{Proposition} 
 
\begin{figure}
 \begin{center}
  \includegraphics[scale=0.8]{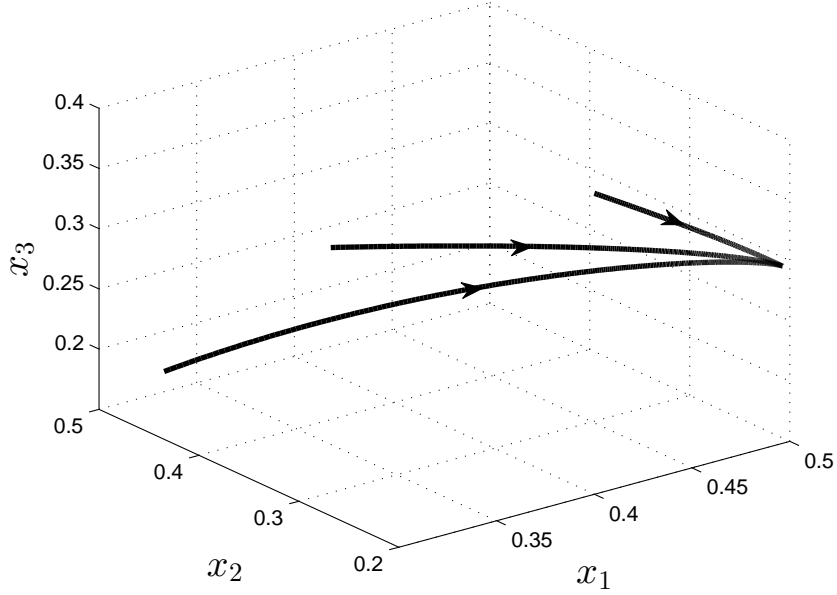}
	\caption{Trajectories of the {\rfmrshort} in
	Example~\ref{exa:n3init} from 
	three different initial conditions in~$L_1$.}
	\label{fig:rfmrd_n3_3initials}
\end{center}
\end{figure}

\begin{Example}\label{exa:n3init}
Consider the {\rfmrshort} with~$n=3$, $\lambda_1=\lambda_2=\lambda_3=1$, $q_1=1$, $q_2=1/2$, and~$q_3=1$.
Fig.~\ref{fig:rfmrd_n3_3initials} depicts the trajectories emanating from three different initial conditions in the level set~$L_1$: $\begin{bmatrix} 
1/3&1/3&1/3\end{bmatrix}^T$, 
$\begin{bmatrix} 2/5&1/5&2/5 \end{bmatrix}^T$,
 and 
$\begin{bmatrix}  1/3&1/2&1/6 \end{bmatrix}^T$. It may be observed that all   three trajectories converge to the same     equilibrium point~$e^1    =
\begin{bmatrix}   0.5&0.207&0.293
\end{bmatrix}^T$ (all numerical values in this paper are to four-digit accuracy).
\end{Example}

It is clear that~$e^s$ satisfies $\sum_{i=1}^{n} e_i^s=s$, and also
\begin{align}\label{eq:ewqqp}
\lambda_n e_n ^ s(q_1-e_1^ s)&=\lambda_1 e_1^ s (q_2-e_2^ s) \\
&=\lambda_2 e_2^ s (q_3-e_3^ s) \nonumber \\
&\vdots  \nonumber \\
&=\lambda_{n-1} e_{n-1}^ s (q_n-e_n^ s).\nonumber 
\end{align}
In other words, at the steady-state the flow into and out of
 each site is equal.
Let
\begin{align*}
R^s:=\lambda_{n-1} e_{n-1}^ s (q_n-e_n^ s)
\end{align*}
denote this  steady-state flow rate for any
 initial condition  in~$L_s$. 

Note that~$L_0$ includes only the origin and for this initial condition~$x(t)\equiv 0$, so~$e^0=0$, and~$R^0=0$. Let~$p:=q_1+\dots+q_n$. Then~$L_p$ includes only the point~$q:=\begin{bmatrix}q_1&\dots&q_n \end{bmatrix}'$ and for this initial condition~$e^p=q$, and~$R^p=0$. Thus, the steady-state flow is zero in both these extreme cases.

\subsection{Optimal steady-state flow  }
A natural question is how does~$R^s$ depends on~$s$? 
When~$s$ is very small 
we expect a small~$R^s$ because there are few particles along the circular chain.
When~$s$ is very large 
we again expect a small~$R^s$ because there are too many particles along the circular chain and this yields
``traffic jams''. The next result shows that there exists
 a unique total density~$s^*$ that maximizes the steady-state flow rate. We refer to this   
as the \emph{optimal density}. 

\begin{Proposition}\label{prop:equi}
Consider an~{\rfmrshort} with rates~$\lambda_i$
and site sizes~$q_i$. There exists a unique
value~$s^*=s^*(\lambda_1,\dots,\lambda_n,q_1,\dots,q_n)  
 \in [0,q_1+\dots+q_n]$
such that~$R^*:=R^{s^*} > R^s$ for any~$s\not =s^*$.		
Furthermore,~$R^s$ is increasing in~$s$ for all~$s<s^*$
and decreasing in~$s$ for all~$s>s^*$. 
Let~$e^*$   denote the
  steady-state corresponding to the 
	density~$s^*$. Then 
\be\label{eq:lgtpp}
			e_1^*e_2^* \dots e_n^*=(q_1-e_1^*)(q_2-e_2^*) \dots (q_n-e_n^*).
\ee
\end{Proposition}

Eq.~\eqref{eq:lgtpp}  can be explained as follows. If $s$ is very small, then 
every~$e_i^s$ is small   (as~$\sum_{i=1}^n e_i^s=s$)  and  
the left-hand side of~\eqref{eq:lgtpp} is smaller than the right-hand side.
 If $s$ is very large, then the opposite case  occurs. The optimal $s^*$ is the value
that yields an equality in~\eqref{eq:lgtpp}.

\begin{figure*}[t]
 \begin{center}
	  \includegraphics[scale=0.8]{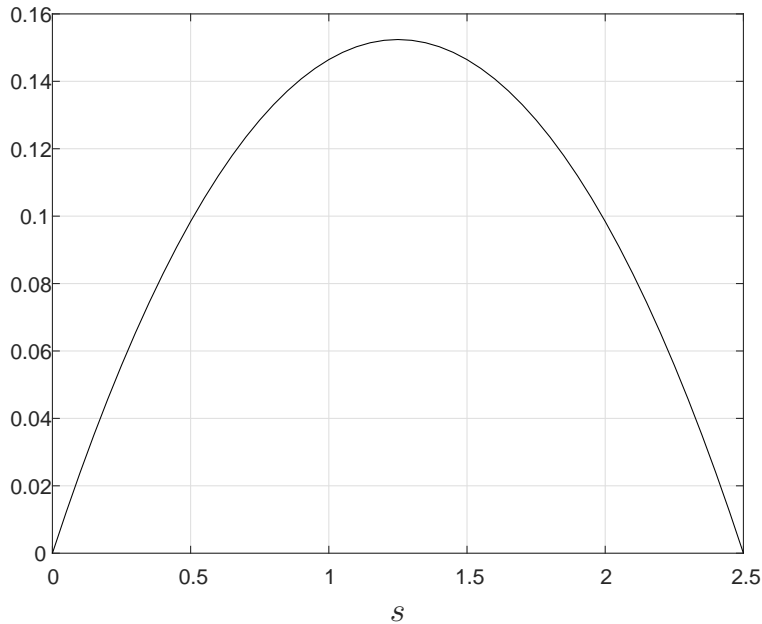}
	\caption{Steady-state flow rate~$R^s$ as a function of $s$ in Example~\ref{exa:mars}.}
	\label{fig:plots}
\end{center}
\end{figure*}

\begin{Example}\label{exa:mars}
Fig.~\ref{fig:plots} depicts the 
steady-state flow rate~$R^s$ as a function of~$s$
for an {\rfmrshort} with~$n=3$, $\lambda_1=\lambda_2=\lambda_3=1$, $q_1=1$, $q_2=1/2$, and~$q_3=1$. 
This was generated by simulating the dynamics until convergence 
for various values of~$s$   with an initial condition~$x(0)$
satisfying~$x_i(0)\in[0,q_i]$ and~$\sum_{i=1}^3 x_i(0)=s$.
The value that maximizes~$R^s$ is~$s^*=1.25$
(i.e. one half of the maximal possible total   density 
which  is~$q_1+q_2+q_3=2.5$), and  the corresponding steady-state is 
\be\label{eqLoptyy}
											e_1^*=0.6096 ,\; e_2^*= 0.2500 ,\; e_3^* = 0.3904.
\ee
A calculation shows that these values satisfy~\eqref{eq:lgtpp}. Note that here site~$2$ is the ``bottleneck site'' in the sense that its size  is smaller than that of the other two sites, and that~$e^*_2=q_2/2$, i.e. the optimal density at site~$2$ is exactly one
half of its capacity.  
\end{Example}

\begin{Example}\label{exa:RFMRD2}
Consider an {\rfmrshort} of order $n=2$, 
\begin{align*}
\dot x_1&=\lambda_2 x_2(q_1-x_1)-\lambda_1 x_1(q_2-x_2),\\
\dot x_2&=\lambda_1 x_1(q_2-x_2)-\lambda_2 x_2(q_1-x_1).
\end{align*}
The steady-state  satisfies  
$
\lambda_2 e_2(q_1-e_1) =\lambda_1 e_1 (q_2-e_2 ),
 $
and this yields
\be\label{eq:expe2456} 
	e_2=\frac{\lambda_1 e_1 q_2} {(\lambda_1-\lambda_2) e_1 +\lambda_2 q_1} .
\ee
The steady-state flow rate is thus
\begin{align*}
R &= \lambda_2 e_2 (q_1-e_1)\\
&= \frac{\lambda_1 \lambda_2   e_1 q_2(q_1-e_1)} {(\lambda_1-\lambda_2) e_1 +\lambda_2 q_1} .
\end{align*}
Differentiating this expression with respect to~$e_1$ and setting the result to zero yields 
two solutions for~$e_1$. The feasible one (i.e. the one in~$[0,q_1]$) is
\[
e_1^*=\frac{  q_1} {1+\sqrt{ \frac{\lambda_1}{\lambda_2} } }.
\]
It is straightforward to verify that this corresponds to a maximum of~$R$. 
Now~\eqref{eq:expe2456} yields
\begin{align*}
e_2^*=\frac{  q_2} {1+\sqrt{ \frac{\lambda_2}{\lambda_1} } },
\end{align*}
and it is straightforward to verify that indeed 
\[
e_1^*e_2^*  =(q_1-e_1^*)(q_2-e_2^*).
\]

Note also that here~$e_i^*=c_i(\lambda_1,\lambda_2)q_i$, with~$c_i\in(0,1)$.
This means that the optimal density at site~$i$ increases with~$q_i$. 
Also, $e_1^*$ decreases and~$e_2^*$ increases when the ratio~$\lambda_1/ \lambda_2$ increases.  This makes sense,
as~$\lambda_1$ controls the exit rate from site~$1$ and the input rate into site~$2$,
whereas~$\lambda_2$ controls the input rate into site~$1$ and the exit rate from site~$2$.
\end{Example}

So far we determined~$e^*$ and~$R^*$  by solving  
   equations~\eqref{eq:ewqqp}
 and~\eqref{eq:lgtpp}. These equations are nonlinear and furthermore they provide little insight 
on the properties of~$e^*,R^*$. It turns out that there is a different and more
useful representation of the optimal steady-state values. 
This representation depends on the Perron root and Perron vector of a 
specific componentwise  nonnegative   matrix.

Given the  {\rfmrshort}~\eqref{eq:rfmrl}, define a parameter-dependent
matrix~$A:\R_+ \to \R^{n\times n}$ by
\be\label{eq:pjm}
A(\kappa ):=\kappa D(q)+B(\lambda),
\ee
where~$D(q)$ is the diagonal matrix with entries
$ 1-q_1,1-q_2,...,1-q_n$ on the diagonal, 
and
\be\label{eq:defmatb}
B(\lambda):=\left [\begin{smallmatrix}
0 &  \lambda_1^{-1/2}   & 0 &0 & \dots &0&\lambda_n^{-1/2} \\
\lambda_1^{-1/2} & 0 & \lambda_2^{-1/2}   & 0  & \dots &0&0 \\
 0& \lambda_2^{-1/2} & 0 &  \lambda_3^{-1/2}    & \dots &0&0 \\
 & &&\vdots \\
 0& 0 & 0 & \dots &\lambda_{n-2}^{-1/2}  & 0 & \lambda_{n-1}^{-1/2}     \\
 \lambda_n^{-1/2}& 0 & 0 & \dots & 0 & \lambda_{n-1 }^{-1/2}  & 0
 \end{smallmatrix}\right ].
\ee
Note that~$B$ is componentwise nonnegative and irreducible. 
Matrices in the form~\eqref{eq:pjm} are sometimes called \emph{periodic Jacobi matrices} (see, e.g.~\cite{pjacomatrix}).
We emphasize that
 the parameters~$q_i$ and~$\lambda_i$ in~$D(q)$ and~$B(\lambda)$
   are the site sizes and transition rates of~\eqref{eq:rfmrl}. 

The matrix~$A(\kappa)$ is componentwise nonnegative and irreducible for all $\kappa\ge 0$ and the Perron-Frobenius theory~\cite{matrx_ana} implies that it admits a simple 
 eigenvalue~$\sigma( \kappa):=\sigma(A(\kappa))$ that is positive and 
larger than the modulus of any other eigenvalue. 
Let~$\zeta( \kappa)\in\R^n_{++}$ denote the corresponding Perron vector,  that is,~$A(\kappa)\zeta(\kappa)=\sigma(\kappa)\zeta(\kappa)$.

\begin{Theorem}\label{thm:pcfv}
Consider the~{\rfmrshort} with~$n>2$. 
There exists a unique value~$ \kappa^*\in [0,\infty)$ such that the matrix~$A(\kappa)$
satisfies
\be\label{eq:gikka}
\sigma(  \kappa^* )=\kappa^* . 
\ee
The optimal steady-state densities~$e^*$ and flow
 rate~$R^*$ of~\eqref{eq:rfmrl} satisfy
\be\label{eq:opfece}
										R^*=	(\sigma (  \kappa^*) )^{-2}=  (\kappa^*) ^{-2},
\ee
and
\be\label{eq:sedopo}
					e^*_i=\frac{\zeta_{i+1}(  \kappa^*)}{\lambda_i^{1/2} \sigma (  \kappa^*)  \zeta_{i}( \kappa^*)} , 
					\quad i=1,\dots,n, 
\ee
(recall that all indexes are interpreted modulo~$n$, so in particular~$\zeta_{n+1}(  \kappa^*)=\zeta_{1}(  \kappa^*)$).
\end{Theorem}

This provides a \emph{spectral representation}
 for~$e^*$ and~$R^*$ in~{\rfmrshort}.
 The proof of Thm.~\ref{thm:pcfv}
(see the Appendix) uses the 
  function 
\be\label{eq:deffk}
f(\kappa):= \sigma (A(\kappa) )-\kappa, 
\ee
and shows  that~$f(0)>0$, $\lim_{\kappa \to \infty } 
f(\kappa)=-\infty$ and~$\frac{d}{d \kappa}f(\kappa)<0$ for all~$\kappa \geq 0$. This implies that there exists a unique value~$\kappa^*$ as described above, and also   
that it is easy to numerically determine~$\kappa^*$ using for example
a simple bisection algorithm. 

Let~$\eta>0$ denote the Perron root of~$B(\lambda)$. 
If~$q_i=1$ for all~$i$ then~\eqref{eq:pjm} gives~$\sigma(A(\kappa))=\eta$
for all~$\kappa$, so the solution of~\eqref{eq:gikka}
is~$\kappa^*=\eta$ and~\eqref{eq:opfece} becomes 
							$			R^*=	\eta^{-2}$.
This recovers the spectral representation of the steady-state 
in the~RFMR~\cite{alexander2017}. Note however that the
spectral representation of the steady-state in the~{\rfmrshort} is 
quite different than the one in the~RFMR as it includes two steps: 
determining the value~$\kappa^*$ and then using the Perron root and Perron vector of~$A(\kappa^*)$.

The next two examples demonstrate Thm.~\ref{thm:pcfv}.

\begin{Example}\label{exa:rfmr3_spectralcomparison}
Consider an {\rfmrshort} with~$n=3$.
Recall that the  optimal steady-state solution satisfies:  
\begin{align}\label{eq:opstre}
\lambda_3 e_3^*(q_1-e^*_1) &=\lambda_1 e^*_1(q_2-e^*_2)=\lambda_2 e^*_2(q_3-e^*_3) \\
e_1^*e_2^* e_3^* &=(q_1-e_1^*)(q_2-e_2^*)(q_3-e_3^*).
\end{align}
For~$\lambda_1=\lambda_2=\lambda_3=1$, $q_1=q_3=1$ and~$q_2=1/2$ the feasible solution of~\eqref{eq:opstre}
(i.e. the solution satisfying~$e^*_i \in [0,q_i]$ for all~$i$) is
\be\label{eq:e1e2e3opt}
e_1^* = \frac{9 - \sqrt{17}}{8}\approx 0.6096 ,\; e_2^* = \frac{1}{4}, \; e_3^* = \frac{-1 + \sqrt{17}}{8}\approx 0.3904.
\ee
The steady-state optimal flow rate  is thus
\begin{align}\label{eq:r681met}
R^* = \lambda_2 e^*_2(q_3-e^*_3) = \frac{1}{32} ( 9-\sqrt{17}    ) \approx 0.1524.
\end{align}

On the other hand, for these parameter values the matrix in~\eqref{eq:pjm}  is 
$
A(\kappa) 
 = \begin{bmatrix}
0 & 1 & 1\\
1&  \kappa/2&  1 \\
1&  1 & 0 
 \end{bmatrix}.
$ 
The Perron root of~$A(\kappa)$  is
$ (2 + \kappa + \sqrt{36 - 4 \kappa + \kappa^2})/4$, so~$\kappa^* \geq 0$ is the solution of
\[
 \kappa =(2 + \kappa + \sqrt{36 - 4 \kappa + \kappa^2})/4
\]
 yielding~$\kappa^*=  (1 + \sqrt{17})/2$.
Thus, Thm.~\ref{thm:pcfv} implies that 
\[
R^*=(\sigma( \kappa^*))^{-2}=((1 + \sqrt{17})/2)^{-2}\approx 0.1524 ,
\]
and this agrees with~\eqref{eq:r681met}. 
The Perron vector of~$A(\kappa^*)$ is 
$
				\zeta( \kappa^*)=\begin{bmatrix}  5+\sqrt{17} & 2(3+\sqrt{17}) & 5+\sqrt{17}  \end{bmatrix}^T ,
$
so Thm.~\ref{thm:pcfv} yields
\begin{align*}
e_1^*&=  \frac{4(3+\sqrt{17})}{(5+\sqrt{17})(1+\sqrt{17}) } \approx 0.6096 ,\\
e_2^*&=  \frac{5+\sqrt{17}}{  (3+\sqrt{17}) (1+\sqrt{17}) } =1/4,\\
e_3^*&=\frac{2}{1+\sqrt{17}}\approx 0.3904,
\end{align*}
and this agrees with~\eqref{eq:e1e2e3opt}.  
\end{Example}

\begin{Example}\label{exa:homog}
Consider the special case of an~{\rfmrshort} with all the~$q_i$'s   equal and denote their
 common value by~$q$. 
Then~$A(\kappa)= (1-q)\kappa I +B(\lambda)$.
Let~$\eta >0$ denote the Perron root of~$B$. 
Then the Perron root of~$A(\kappa)$ is~$\sigma( \kappa )=(1-q)\kappa +\eta$,
so the equation~$\sigma( \kappa )= \kappa$ becomes 
$(1-q) \kappa+\eta= \kappa $ and this admits a unique solution
\be
 \kappa^*=\eta/q.
\ee
 The Perron
 vector~$\zeta^*:=\zeta (k ^*)$ of~$A^*:=A(\kappa^*)$
 satisfies
$
							A^* \zeta^* =\kappa^* \zeta^*
$
 and this gives~$B\zeta^* =\eta \zeta^* $. Thus,~$\zeta^*$ is the Perron  vector of~$B$. 
If, in addition, all the~$\lambda_i$'s are equal, with~$\lambda$ denoting their common value, then it is straightforward to verify that the Perron root and  vector of~$B$
are~$\eta=2\lambda^{-1/2}$ and~$\mu:=\begin{bmatrix}1&\dots &1\end{bmatrix}^T$. We conclude that if~$q_1=\dots=q_n=q$
and~$\lambda_1=\dots=\lambda_n=\lambda$ then
$\kappa^*=2\lambda^{-1/2}/q$, 
and~$\zeta^*=\begin{bmatrix}1&\dots &1\end{bmatrix}^T$,
so the spectral representation yields 
\be\label{eq:pofs}
  e_i^*=  q/2 \text{ for all }i, \text{ and }  R^*= q^2 \lambda/4.
\ee
   Note that in this case the optimal steady-state density and flow rate do not depend on~$n$ (yet the optimal total density~$s^*$ does depend on~$n$, as~$s^*=\sum_{i=1}^n q/2=qn/2$).
	It is important to note that~\eqref{eq:pofs} shows that~$\lambda$ and~$q$
	play a very different role in determining~$R^*$. In particular,
	a small value of~$q\in(0,1]$ will decrease~$R^*$ more than a small value of~$\lambda$. 
\end{Example}

It is intuitively clear that even if one of the
 rates in the~{\rfmrshort}
goes to infinity  the densities and production rate  remain bounded, as the other rates constrain the dynamics. The next result states this formally for the  optimal density case. As we will see below this will prove useful
 in analyzing  the~{\rfmshort}.

\begin{Corollary}\label{coro:vobu}
The optimal-density 
production rate and densities 
in the~{\rfmrshort} remain bounded if~$\lambda_i \to \infty$ for some~$i$. 
\end{Corollary}

The spectral representation of the optimal steady-state
 in the~{\rfmrshort} 
has important theoretical and practical
 implications. Two of these are discussed in the remainder of this section. 

\subsection{Sensitivity Analysis}
For any model that admits 
  a steady-state a
natural and important question is: suppose that we make a small change in one of the parameters, what is the resulting
change in the steady-state values? 
For the~RFM, this kind of
 sensitivity analysis  has 
 appeared in~\cite{RFM_sense}. Here, we use the spectral representation
 to analyze  the sensitivity
 of  the optimal-density
 steady-state flow rate in the~{\rfmrshort}.

Consider an {\rfmrshort} with dimension~$n$. 
Let~$p:=\begin{bmatrix} \lambda_1 &\dots& \lambda_n & q_1& \dots& q_n \end{bmatrix}^T$ denote its set of parameters, with~$p \in \R^n_{++}\times (0,1]^n$. We know that~$p$ induces an optimal density~$s^*$ and that for any initial condition~$a\in[0,q_1]\times\dots\times[0,q_n]$,
with~$\sum_{i=1}^n x_i(0)=s^*$, 
the solution~$x(t,a)$ converges to a steady-state 
density~$e^*=e^*(p)$ and flow rate~$R^*=R^*(p)$. 
These steady-state values can be obtained from the spectral representation described in 
Thm.~\ref{thm:pcfv}.  
\begin{Proposition} \label{prop:Rsens_lambda_and_q}
Consider an 
{\rfmrshort} with dimension~$n$.
Let~$\kappa^*\geq 0$
denote  the unique solution of~$\sigma(A(\kappa))=\kappa$, and let~$\zeta^* \in \R^n_{++}$
denote the Perron  vector of~$A(\kappa^*)$
normalized such that~$(\zeta^*)^T\zeta^*=1$.
For any~$i\in\{1,\dots,n\}$ 
the  sensitivity of~$R^*$  with respect to a change of parameters  is given by
\begin{align}\label{eq:senselam}
\frac{d}{d \lambda_i } R^*  &= 2 \left( 1- \sum_{i=1}^n (1-q_i)(\zeta^*_i)^2  \right ) ^{-1}  \zeta^*_i \zeta^*_{i+1} \lambda_i^{-3/2} (k^*)^{-3},
\end{align}
and
\begin{align}\label{eq:dfvcf}
\frac{d}{d q_i } R^* &= 2  \left( 1- \sum_{i=1}^n (1-q_i)(\zeta^*_i)^2  \right ) ^{-1}      (\zeta^*_i)^2   (k^*)^{-2}. 
\end{align}
\end{Proposition}

\begin{Remark}\label{rem:posmon}
Note that since~$q_i \in (0,1]$,~$\zeta^*_i>0$, and~$\sum_{i=1}^n (\zeta^*_i)^2=1$, this implies that~$\frac{d}{d \lambda_i } R^* >0$
and~$\frac{d}{d q_i } R^* >0$, that is,
an increase [decrease]
in any transition rate or site size 
increases [decreases] 
the optimal steady-state flow rate. 
This makes sense, as increasing $\lambda_i$  
increases the flow rate from site~$i$ to site~$i+1$ whereas increasing $q_i$ increases  the  capacity at site~$i$, and both 
improve the flow rate and decrease ``traffic jams''.
\end{Remark}



\begin{Example}
Consider again
the {\rfmrshort} 
with $n=3$   and parameters~$\lambda_i=1,i=1,2,3, q_1=q_3=1,q_2=1/2$.
Recall from  Example~\ref{exa:rfmr3_spectralcomparison} 
that in this case 
the Perron root of~$A(\kappa^*)$ is:
\[
\sigma (A(\kappa^*)) = \kappa^* = (1+\sqrt{17})/2 ,
\]
and the optimal steady-state flow rate is thus
\be
										R^*=	(\sigma (A( \kappa^*) ))^{-2} = (9-\sqrt{17})/32 .
\ee
The corresponding normalized Perron  vector
is 
\[
				\zeta ^*=\frac{\begin{bmatrix}  5+\sqrt{17} & 2(3+\sqrt{17}) & 5+\sqrt{17}  \end{bmatrix}^T}{ \sqrt{188+44\sqrt{17}}}.
\] 
Calculating the sensitivity with respect to~$\lambda_2$   using~\eqref{eq:senselam}
yields
\begin{align}\label{eq:latpao}
\frac{d}{d \lambda_2 } R^*  &= 2 ( 1- \sum_{i=1}^n (1-q_i)(\zeta^*_i)^2  ) ^{-1} \zeta_2^* \zeta_3^* \lambda_2^{-3/2} (k^*)^{-3}
\nonumber \\
&= \frac{2 \zeta_2^* \zeta_3^*   (k^*)^{-3}}{ 1 - (1/2)(\zeta^*_2)^2 } \nonumber \\
&= \frac{ 4 (3+\sqrt{17})(5+\sqrt{17})( \frac{1+\sqrt{17}}{2}   ) ^{-3}
   } { 188+44\sqrt{17} -2(3+\sqrt{17})^2 }
	\nonumber \\
&= 0.0577 . 
\end{align}

Let~$\varepsilon:=-0.01$ and suppose that 
 $\lambda_2$ is decreased to~$\overline { \lambda_2} :=\lambda_2+\varepsilon= 0.99 $. A direct calculation of the   optimal steady-state 
flow ratet in the modified~{\rfmrshort}  yields~$\overline  {R^*}=  0.151823$,
so
\[
\frac{\overline  {R^*}-R^*}{\varepsilon}= 0.0580 ,
\]
and this agrees well with~\eqref{eq:latpao}.
\end{Example}

\begin{Example}
   Example~\ref{exa:homog} showed that for 
  an~{\rfmrshort} 
with~$\lambda_1=\dots=\lambda_n=\lambda$ and~$q_1=\dots=q_n=q$ we have~$k^*=2\lambda^{-1/2}/q $,
$R^*= q^2 \lambda/4  $, and the normalized
Perron  vector is
$
\zeta^*=\frac{1}{\sqrt{n}}\begin{bmatrix}
1&\dots&1\end{bmatrix}^T$.
Substituting these values in~\eqref{eq:senselam}
and~\eqref{eq:dfvcf} yields
\begin{align*} 
\frac{d}{d \lambda_i } R^*   =\frac{q^2}{4n},
\text{ and }
\frac{d}{d q_i } R^*  = \frac{\lambda q }{2n}. 
\end{align*}
These results show that although~$R^*$ does not depend on~$n$, the sensitivities decay like~$1/n$. Furthermore, they highlight the different roles of the rates and the site sizes. 
\end{Example}

\subsection{Optimizing the production rate with respect to the site sizes and transition rates }

Let~$E:=\R^n_{++}\times (0,1]^n $. 
We already know that 
any set of parameters~$p:=(\LMDR,\QR)\in E$
 induces an optimal total density~$s^*$,
 and  that the~{\rfmrshort} initialized with this total density
yields a maximal production rate~$R^*$ (with respect to all other initial conditions). 
This yields a mapping~$p \to R^*(p)$. 

 Suppose that we are given a compact subset~$\Omega\subset  E$. 
Every vector in~$\Omega$ can be used as a set of rates and site sizes in
 the~{\rfmrshort}.
A natural goal  is to  determine a vector~$\eta \in \Omega $ that yields the maximal flow rate, that is,
\be\label{eq:esspo}
				R(\eta) = \max_{p \in \Omega} R^*( p).
\ee

In the context of translation, this means that the circular mRNA with parameters~$\eta$, 
  initialized with total density~$s^*(\eta)$, will yield a steady-state production rate that is higher or equal 
	to   that obtained for all the other parameter vectors in~$\Omega$ and all other initial conditions.  

The next result is essential for   analyzing the maximization problem in~\eqref{eq:esspo}.
\begin{Theorem}\label{thm:R^(-1/2)_convex}
The function ${R^*} = {R^*}(\lambda_1,...,\lambda_n,q_1,...,q_n) $ is  quasi-concave over~$E$, that is, for any~$p,\tilde  p\in  E$   we have
\be\label{eq:rcomnac}
	R^*(	r p+(1-r) \tilde p )	 
			\geq \min\{ R^*(p), R^*(\tilde p) \}, \text{ for all } r\in[0,1].
\ee
Furthermore, for fixed~$q_i$'s the function 
${R^*} = {R^*}(\lambda_1,...,\lambda_n )$
is concave over~$\R_{++}^{n} $.
\end{Theorem}

\begin{Example}\label{exa:quasi2}
Consider an {\rfmrshort} with~$n=2$.
We know from  
Example~\ref{exa:RFMRD2} that
the optimal     
  steady-state flow rate is  
\begin{align*}
R^*(p) &=f(\lambda_1,\lambda_2)g(q_1,q_2),
\end{align*}
with~$f(\lambda_1,\lambda_2):=\frac{\lambda_1\lambda_2  }
{  (\sqrt{ \lambda_1}+\sqrt{\lambda_2} )^2}$
 and~$g(q_1,q_2):=q_1q_2$.
In general,~$R^*$ is not convex nor concave. Indeed, 
for~$\lambda_1=\lambda_2=4$,
$
R^*(q_1,q_2)=q_1q_2, 
$
and computing the Hessian of this function shows that it is not convex nor concave. 
However,~$\log(g(q_1,q_2))=\log(q_1)+\log(q_2)$ and this is concave, so~$g(q_1,q_2)$
is log-concave, and thus quasi-concave.
Analysis of the Hessian of~$(-f(\lambda_1,\lambda_2))$ shows that it is 
convex over~$\R^2_{++}$, so~$f$ is concave (and thus log-concave) over~$\R^2_{++}$.
We conclude that the product~$R^*=fg$ is log-concave and thus quasi-concave
over~$\R^2_{++}\times (0,1]^2$. 
\end{Example}
 
The next result 
is an  immediate implication of Thm.~\ref{thm:R^(-1/2)_convex}. 
\begin{Corollary}
Fix a convex set~$\Omega \subseteq \R^n_{++}\times (0,1]^n$.
The problem of maximizing~$R^*(p)$ over~$p\in\Omega $  
is a quasi-concave optimization problem. 
Furthermore, for a fixed set of~$q_i$'s
the problem of maximizing~$R^*(\LMDR)$ over a convex set of~$ \R^n_{++}$  
is a  concave optimization problem. 
\end{Corollary}

An example of such an optimization problem is the following.  
\begin{Problem}\label{prob:const_opt}
Consider an~{\rfmrshort} with dimension $n$. Given~$w_1,\dots,w_{n}, v_1,\dots,v_{n},b>0$, 
\begin{align*}
&\text{Maximize } R^*=R^*(\LMDR,\QR) \\
&\text{ with respect to } \LMDR,\QR
\end{align*}
 subject to the constraints~$\lambda_i>0$, $q_i\in (0,1]$, and 
\begin{align*} 
&\sum_{i=1}^n w_i \lambda_i +\sum_{i=1}^n v_i 
q_i \le b.
\end{align*}
\end{Problem}
In other words, the problem is to maximize $R^*$ w.r.t. the rates $\LMDR$ and site sizes~$\QR$, under the constraint that 
a weighted sum of all the parameters
is bounded by~$b$. The weights~$w_i,v_i$, $i=1,\dots,n$,
 can be used to provide different weighting to the different rates and site sizes, respectively, and $b$ represents a kind of ``total biocellular budget''.
 
\begin{Example}
Consider Problem~\ref{prob:const_opt}
with~$v_i=w_i=1$ for all~$i\in [1,\hdots,n]$ and~$b=n$. 
Thus, the problem is to maximize~$R^*$
 subject to the constraints~$\lambda_i>0$, $q_i\in (0,1]$, and 
\begin{align*} 
&\sum_{i=1}^n   \lambda_i +\sum_{i=1}^n  
q_i \le n.
\end{align*}
By symmetry, there exist~$q,\lambda$ such that
the  solution satisfies~$q_i=q $  and~$\lambda_i =\lambda $ for all~$i$.
Example~\ref{exa:homog} implies that~$R^*= q ^2 \lambda /4  $, so the problem is 
$\max   (q^2 \lambda /4)$ 
 subject to the constraints~$\lambda >0$, $q \in (0,1]$, and 
$   \lambda +q  \le 1 $.
By Remark~\ref{rem:posmon},
  the solution must satisfy~$ \lambda  +q   = 1 $, so the problem is 
$\max   (q^2 ( 1-q) /4)$ 
 subject to~$q \in (0,1]$. It is straightforward to verify  that the  optimal solution is 
\[
		q^*=\frac{2}{3 }, \; \lambda^*=\frac{1}{3 },
\]
yielding~$
R^*=\frac{1}{27 }$.
In other words, the optimal solution is to
 allocate~$2/3$ of the total budget 
 on the~$q_i$'s
and~$1/3$ on the~$\lambda_i$'s. 
\end{Example}

Note again that this highlights the different roles of the 
rates and site sizes.  In the context  of maximizing the 
optimal-density    steady-state flow rate
 the site sizes  are more important than the rates. 

\subsection{Entrainment}
Biological organisms are exposed to periodic
 excitations like the electric impulses produced by the sinoatrial node, the 24h solar day,
and the periodic cell-cycle division program. Proper functioning often  requires internal processes to \emph{entrain} to these excitations, that is, 
to vary periodically with the same period as the excitation. 
There is a considerable interest  
 in understanding the molecular and genetic 
mechanisms  underlying entrainment. Indeed, 
the 2017 Nobel Prize in Physiology or Medicine
was awarded 
to
Jeffrey C. Hall, Michael Rosbash and Michael W. Young
for their discoveries of molecular mechanisms controlling the circadian rhythm.

It is reasonable  to assume   
that protein synthesis is regulated in accordance with the periodic cell-cycle division process.
Indeed, several papers reported that during mitosis
   global translation   is inhibited
at the level of 5'cap-dependent initiation
and also at the level of elongation,
see the review~\cite{regulation_cell_cycle}.
A natural question is whether periodically-varying
patterns of initiation and/or  elongation factors yield 
a periodic pattern of ribosome density and thus a  periodic protein production rate?

In the context of the~{\rfmrshort}, this question of  entrainment can be studied rigorously.  
Suppose   that the transition rates~$\lambda_i$
 along the cyclic chain are not constants, but periodically time-varying functions of time with a
 common (minimal) period~$T>0$. 
In this setting entrainment means that the 
 site densities (and thus production rate)
converge to a periodically varying pattern with the same period~$T$. Note that although this may seem immediate, it is not necessarily so. 
For example, Ref.~\cite{NIKOLAEV20181232} provides examples of low-dimensional and 
``innocent-looking'' nonlinear systems where in response to a periodic excitation   chaotic trajectories arise.

  A function~$f:\R\to\R$ is called~$T$-periodic if $f(t+T)=f(t)$ for all~$t$. Assume that all
	the~$\lambda_i$'s in the~{\rfmrshort} 
	are time-varying with
	\[
	0<\delta_1\leq \lambda_i(t) \leq \delta_2
	\]
	for all~$i$ and all~$t$ and that they are all~$T$-periodic. 
	We refer to the model in this case as the 
	\emph{periodic ribosome flow model on a ring with
	different cell sizes}~({\prfmrshort}). 

\begin{Theorem}  \label{thm:period}
 Consider the~{\prfmrshort}. Fix an arbitrary~$s \in [0,q_1+\dots+q_n]$. There exists a unique function~$\phi^s:\R_+ \to C  $, that is~$T$-periodic,
and
\[
      \lim_{t\to \infty }  |x(t,a)-\phi^s(t) | =0,  \text{ for all }a \in L_s.
\]
\end{Theorem}

In other words, every level set~$L_s$ of~$H$
contains a unique~$T$-periodic solution, and every
trajectory
 of the~{\prfmrshort} with an initial  total density in~$L_s$  converges to this solution.
Thus, the~{\prfmrshort} entrains  to the periodic excitation in the~$\lambda_i$'s.

Note that since a constant function is a~$T$-periodic function for any~$T$,
Thm.~\ref{thm:period} implies entrainment to a
periodic trajectory in the particular case where one of the~$\lambda_i$'s oscillates, and all the other rates are constant.
Note also that the stability part in Prop.~\ref{prop:sta_inv} is a special case of 
Thm.~\ref{thm:period}.

\begin{figure}[t]
  \begin{center}
  \includegraphics[scale=0.6]{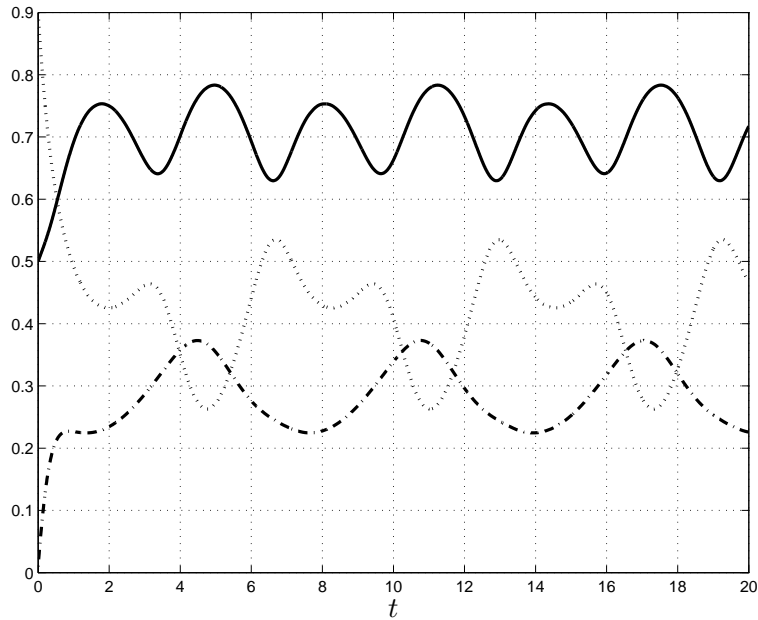}
  \caption{Densities~$x_1(t)$ (solid line),
	$x_2(t)$ (dash-dotted line),
	and
	$x_3(t)$ (dotted line) as a function of~$t$ 
	 in Example~\ref{exa:period}.
  }  \label{fig:period}
  \end{center}
\end{figure}

\begin{Example}\label{exa:period}
Consider the~{\prfmrshort} with~$n=3$, $\lambda_1(t) =3$, $\lambda_2(t)=3+2\sin(t+1/2)$,  $\lambda_3(t)=4-2\cos(2t)$, and 
site sizes~$q_1=q_3=1$ and~$q_2=0.5$.
Note that all the~$\lambda_i$'s are periodic with a (minimal) common period~$T=2 \pi $.
Fig.~\ref{fig:period} shows the solution~$x(t,a)$ for~$a=\begin{bmatrix}  0.50& 0.01 & 0.90   \end{bmatrix}^T$.
It may be seen that every~$x_i(t)$ converges to a periodic function with period~$2\pi$. 
\end{Example}

We now turn to analyze 
the~{\rfmshort}~\eqref{eq:rfmrd}.
\section{Analysis of the {\rfmshort}} \label{sec:main_open}
Our first result 
describes the 
 asymptotic behavior  of the~{\rfmshort}.
 \begin{Proposition}\label{prop:rfmd_sta_inv}
Consider an~{\rfmshort} of dimension~$n$.
The set~$C$ is an invariant set of~\eqref{eq:rfmrd}, and
there exists a unique~$e\in \Int(C)$ such that
\[
					\lim_{t\to \infty} x(t,a)=e \text{ for all } a\in C.
\]
\end{Proposition} 
In other words, the rates and site sizes in the~{\rfmshort} determine a unique steady-state 
 in~$C$, and the solution emanating from any initial condition in~$C$ converges to~$e$.

The steady-state~$e$   of 
the~{\rfmshort}  can be obtained from that of a
higher-dimensional 
optimal-density~{\rfmrshort}.
We begin with a simple example demonstrating this. 

\begin{Example}
Consider an~{\rfmrshort} with~$n=4$
\begin{align*}
\dot x_1&= \lambda_4 x_4(q_1-x_1)  -\lambda_1 x_1(q_2-x_2) ,\nonumber \\
\dot x_2&=  \lambda_1 x_1(q_2-x_2)-\lambda_2 x_2(q_3-x_3), \nonumber\\
\dot x_3&=   \lambda_2 x_2(q_3-x_3) - \lambda_3 x_3(q_4-x_4), \nonumber\\
\dot x_4&=   \lambda_3 x_3(q_4-x_4) - \lambda_4 x_4(q_1-x_1) .
\end{align*}
Assume that this is initialized with an initial condition corresponding to 
the optimal density, so that the steady-state satisfies
\be\label{eq:spor}
e_1^*e_2^*e_3^*e_4^*=(q_1-e_1^*)(q_2-e_2^*)(q_3-e_3^*)(q_4-e_4^*),
\ee
and
\begin{align}\label{eq:loimp4}
 \lambda_4 e_4^*(q_1-e_1^*)  &=\lambda_1 e_1^*(q_2-e_2^*) \nonumber \\
    &=\lambda_2 e^*_2(q_3-e^*_3) \nonumber \\
		&=\lambda_3 e^*_3(q_4-e^*_4)  .
\end{align}
Suppose that we fix~$\lambda_1,\lambda_2,\lambda_3\in\R_{++}$, $q_1,q_2,q_3,q_4\in(0,1]$,
 and take~$\lambda_4 \to\infty$. 
Then~\eqref{eq:loimp4} suggests  that~$ e_4^*(q_1-e_1^*) \to 0$. 
As we will show in the proof of Prop.~\ref{prop:rfmrd_to_rfmd}
below, we actually have
\be\label{eq:lome1e4}
e_4^*\to 0 \text{ and } e^*_1 \to q_1. 
\ee
 Intuitively, this can be  explained as follows. As~$\lambda_4\to\infty$ the exit rate from site~$4$ is very large, so this site is   emptied i.e.~$e_4^*\to 0 $. Also, the input rate to site~$1$ is very large, and this yields~$e^*_1 \to q_1$ (but the last argument is in fact valid only in 
 the optimal-density~{\rfmshort}).
 Substituting~\eqref{eq:lome1e4}
in~\eqref{eq:loimp4}
implies that when~$\lambda_4\to \infty$, 
\begin{align}\label{eq:limytpr4}
 \lambda_1 q_1 (q_2-e_2^*)  
    &=\lambda_2 e^*_2 (q_3-e_3^*) \nonumber \\
		&=\lambda_3 e^*_3 q_4   .
\end{align}

Now consider an~{\rfmshort} with~$n=2$, 
rates 
$
 [\tilde \lambda_0,\tilde \lambda_1,\tilde \lambda_2]:=[\lambda_1 q_1 ,\lambda_2 ,\lambda_3 q_4]
$,
and site sizes
$
 [\tilde q_1,\tilde q_2]:=[q_2,q_3],
$
 that is,  the system  
\begin{align*}
\dot {\tilde x}_1&=(\lambda_1 q_1)(q_2-\tilde x_1)- \lambda_2 \tilde x_1  (q_3-\tilde x_2) , \\
\dot {\tilde x}_2&= \lambda_2 \tilde x_1 (q_3-\tilde x_2)-(\lambda_3 q_4) \tilde x_2.
\end{align*}
The steady-state~$\tilde e=\begin{bmatrix} \tilde e_1 & \tilde e_2
\end{bmatrix}^T$ of this~{\rfmshort} 
satisfies
\begin{align*}
\lambda_1 q_1 (q_2- \tilde e_1)&=\lambda_2 \tilde e_1 (q_3- \tilde e_2)  \\
&= \lambda_3 q_4 \tilde e_2.
\end{align*}
Comparing this with~\eqref{eq:limytpr4} we conclude that
\begin{align*}
					\tilde e_1=e_2^*,\quad 
					\tilde e_2=e_3^*.
\end{align*}
Thus, we can analyze the steady-state
of a two-dimensional~{\rfmshort} using the results already derived for a
 four-dimensional optimal-density~{\rfmrshort}
 and taking~$\lambda_4\to\infty$.
\end{Example}

The same behavior 
 holds for any dimension. If we take  
   an~{\rfmrshort} with dimension~$n+2$, initialized with the optimal density, and take~$\lambda_{n+2} \to \infty$ then~$e_1^*\to q_1$ and~$e_{n+2}^*\to 0$. 
	This means that site~$1$ [$n+2$] becomes a full [empty]
	reservoir, and   sites~$2,3,\dots,n+1$ in between
	become an open chain 
	that is fed by [feeding] the full [empty]
	reservoir, i.e. an~{\rfmshort}.
	
\begin{Proposition}\label{prop:rfmrd_to_rfmd}
Let~$e^*=\begin{bmatrix}  e^*_1&\dots&e^*_{n+2} \end{bmatrix}^T$ denote the optimal-density
 steady-state of an~{\rfmrshort} with 
dimension~$n+2$,
 rates~$\LMDRR$, and site sizes~$\QRR$.
Let~$\tilde e=\begin{bmatrix} \tilde e_1&\dots&\tilde e_n \end{bmatrix}^T$ denote the steady-state
of an~{\rfmshort} with dimension~$n$,
 rates
\begin{align}\label{eq:addrates}
&\begin{bmatrix}
\tilde \lambda_0&\tilde \lambda_1 & \tilde \lambda_2&\dots& \tilde \lambda_{n-1} & \tilde \lambda_n \end{bmatrix}\nonumber \\
&:=\begin{bmatrix}
\lambda_1 q_1 &\lambda_2  &\lambda_3  &\dots& 
 \lambda_{n}  
& \lambda_{n+1} q_{n+2} \end{bmatrix} , 
\end{align}
and
site sizes
\be\label{eq:addqia}
  \begin{bmatrix}
\tilde q_1&\tilde q_2&\dots& \tilde q_n \end{bmatrix}
:=\begin{bmatrix}
q_2&q_3&\dots& q_{n+1} \end{bmatrix}.
\ee
Then
\be\label{eq:propr77}
\tilde e 
= \lim_{\lambda_{n+2}\to \infty} \begin{bmatrix} 
e_2^*&e_3^*&\dots& e_{n+1}^* \end{bmatrix}^T.
\ee
\end{Proposition}

Thus, we can obtain~$\tilde e$ in the~{\rfmshort}
from the optimal-density steady-state~$ e^*$
in the~{\rfmrshort}.

The next example demonstrates Prop.~\ref{prop:rfmrd_to_rfmd}.

\begin{Example}
Consider an~{\rfmrshort} with dimension $n=5$, and parameters
$\lambda_1=q_1=0.8,\lambda_2=q_2=0.6,\lambda_3=q_3=0.4,\lambda_4=q_4=0.7$, and $\lambda_5=q_5=0.5$. 
The optimal total initial density and  steady-state values are:
\begin{align*}
s^* &= 1.5,\; 
e^* =\left [\begin{smallmatrix} 0.3625 & 0.4448 & 0.2364 & 0.2343 & 0.2170 \end{smallmatrix} \right]^T, R^*  =  0.04748.
\end{align*}
For $\lambda_5=100$  the   values are:
\[
s^* =1.63 , \;
e^*=\left [ \begin{smallmatrix} 0.7246 & 0.5123 & 0.2344 & 0.1517 & 0.0069 \end{smallmatrix}\right]^T,  R^*  =0.0523,
\]
and for $\lambda_5=200$  they are:
\begin{align}\label{eq:exp_l5large}
s^* &=1.65, \;
e^*=\left [ \begin{smallmatrix} 0.7445 & 0.5146 & 0.2348 & 0.1514 & 0.0047 \end{smallmatrix}\right]^T,  R^* =0.0524.
\end{align}
It may be seen 
 that as $\lambda_5$ increases  the optimal-density
 steady-state   at site~$1$
  [site~$5$] increases
[decreases] to $q_1$ [$0$]. 

On the other hand, for an~{\rfmshort} with dimension $n=3$,
 rates 
\begin{align*}
\tilde\lambda&= 
\begin{bmatrix}  \lambda_1 q_1&\lambda_2&\lambda_3&\lambda_4 q_5 \end{bmatrix}^T \\
&=\begin{bmatrix}  0.64 & 0.3 & 0.4 &0.35 \end{bmatrix}^T
\end{align*}
and site sizes
\begin{align*}
\tilde q&=\begin{bmatrix}  q_2&q_3&q_4  \end{bmatrix}^T\\
 &=\begin{bmatrix} 0.6 & 0.4 & 0.7  \end{bmatrix}^T
\end{align*}
 the steady-state values are~$\tilde e=\begin{bmatrix}0.5195 & 0.2345 & 0.1485 \end{bmatrix}^T$, and $\tilde R=0.052$ (compare   with~\eqref{eq:exp_l5large}).
\end{Example}

 Prop.~\ref{prop:rfmrd_to_rfmd}
shows how to reduce an~$(n+2)$-dimensional {\rfmrshort} 
into an~$n$-dimensional~{\rfmshort}.
The next remark shows how we can use this construction in the opposite direction.

\begin{Remark}\label{rem:copos}
Given an~$n$-dimensional~{\rfmshort} with rates~$\tilde\lambda \in \R^{n+1}_{++}$
 and site sizes~$\tilde q \in (0,1]^n$, let~$\tilde e \in  C$ denote its steady-state. 
Define an~$(n+2)$-dimensional {\rfmrshort} with rates
\be\label{eq:rem2rates}
\lambda:=\begin{bmatrix} \tilde \lambda_0 &\tilde  \lambda_1&\dots& 
\tilde \lambda_n & a
\end{bmatrix}^T,
\ee
where~$a>0$, 
and site sizes
\be\label{eq:rem2sir}
q:=\begin{bmatrix}1&\tilde q_1&\tilde q_2&\dots& \tilde q_n & 1 \end{bmatrix}^T.
\ee
Let~$e^*(a) $ denote the optimal-density
steady-state of this~{\rfmrshort}.
Then Prop.~\ref{prop:rfmrd_to_rfmd} implies that
\be\label{eq:ealimi}
\lim_{a\to \infty} e^*(a)=\begin{bmatrix}
1& \tilde e_1 &\tilde e_2&\dots& \tilde e_n &0 
\end{bmatrix}^T.
\ee
\end{Remark}

Using the connection between the optimal-density~{\rfmrshort}
and the~{\rfmshort} we can extend many of the 
 analysis  results derived above for the~{\rfmrshort} to the~{\rfmshort}.
The next result provides a spectral representation for steady-state of the~{\rfmshort}.
\begin{Corollary}\label{cor:rfmd_spectral}
Given an~$n$-dimensional~{\rfmshort} with rates~$\tilde\lambda \in \R^{n+1}_{++}$
 and site sizes~$\tilde q \in (0,1]^n$, let~$\tilde e \in  C$ denote its steady-state. 
Define~$\tilde A:\R_+\to\R^{(n+2)\times(n+2)}$ by
\be\label{eq:atildd}
\tilde A(\kappa):=\left [ \begin{smallmatrix}
0 & \tilde \lambda_0^{-1/2} & 0 & 0 & \dots & 0 & 0 \\
\tilde  \lambda_0^{-1/2} & (1-\tilde  q_1)\kappa &\tilde  \lambda_1^{-1/2} & 0 & \dots & 0 & 0 \\
0 & \tilde  \lambda_1^{-1/2} & (1-\tilde  q_2)\kappa & \tilde  \lambda_2^{-1/2} & \dots & 0 & 0 \\
 &&&\vdots \\
0 & 0 & 0 & \dots &\tilde  \lambda_{n-1}^{-1/2}  & (1-\tilde q_{n})\kappa &\tilde  \lambda_{n}^{-1/2} \\
0 & 0 & 0 & \dots & 0 & \tilde  \lambda_{n }^{-1/2}  & 0
 \end{smallmatrix} \right ].
\ee
Then there exists a unique value~$\tilde  \kappa^*\in [0,\infty)$ such that 
\be\label{eq:gikka_rfm}
\sigma(\tilde A(\tilde  \kappa^*))= \tilde \kappa^* . 
\ee
Let~$\tilde  \zeta  \in\R^{n+2}_{++}   $ denote the Perron 
 vector of~$\tilde A(\tilde \kappa^*)$.
The   steady-state flow rate and densities
in the~{\rfmshort} satisfy
\be\label{eq:opfece_rfm}
		\tilde R=	(\sigma ( \tilde A (\tilde  \kappa^*) ) )^{-2},
\ee
and
\be\label{eq:xsaxo}
					\tilde e_i=\frac{\tilde \zeta_{i+2} }{\tilde \lambda_{i }^{1/2}  \tilde  \kappa^*  \tilde  \zeta_{i+1} } , 
					\quad i=1,\dots,n .  
\ee
\end{Corollary}

\begin{Example}\label{exa:rfmd_n1}
Consider an {\rfmshort} of order $n=1$, 
\begin{align*}
\dot {\tilde x}_1&=\tilde \lambda_0 (\tilde q_1-\tilde x_1)-\tilde \lambda_1 \tilde x_1.
\end{align*}
The steady-state  
satisfies~$\tilde \lambda_0 (\tilde q_1-\tilde e_1)=\tilde \lambda_1 \tilde e_1$, that is,  
\be\label{eq:ertyuo}
\tilde e_1= \frac{\tilde \lambda_0 \tilde q_1 }{\tilde \lambda_0+\tilde \lambda_1}, 
 \ee
and this yields
\be\label{eq:expeRR} 
	\tilde R= \tilde\lambda_1\tilde e_1= \frac{\tilde \lambda_0 \tilde \lambda_1 \tilde q_1 }{\tilde \lambda_0+\tilde \lambda_1}.
\ee

In this case, the spectral representation is based on the matrix  
\[
\tilde A(\kappa) =\begin{bmatrix}
0 & \tilde \lambda_0^{-1/2} & 0  \\
\tilde  \lambda_0^{-1/2} & (1-\tilde  q_1)\kappa &\tilde  \lambda_1^{-1/2}   \\
  0 & \tilde  \lambda_{1 }^{-1/2}  & 0
 \end{bmatrix}.
\]
The Perron root of this matrix is
$
\sigma(\kappa)= \frac{1}{2}\left( 
(1-\tilde q_1)\kappa+
\sqrt { \frac{  4\tilde \lambda_1+ (   4+\kappa^2\tilde\lambda_1 (\tilde q_1-1)^2  )  \tilde \lambda_0      }{ 
\tilde \lambda_0 \tilde\lambda_1  }}
\right ),
$
and thus the 
 unique positive solution of~$\sigma(\kappa)= \kappa$ is
$
\tilde \kappa^*=\sqrt { \frac{   \tilde \lambda_0+
\tilde \lambda_1  }{ 
\tilde \lambda_0 \tilde\lambda_1  \tilde q_1 }}      . 
$
The Perron vector of~$A(\tilde  \kappa^*)$ is
$
\tilde  \zeta =\begin{bmatrix}
\sqrt{ \frac{\tilde \lambda_1 }{\tilde \lambda_0} }  &
\sqrt{ \frac{\tilde \lambda_0+\tilde \lambda_1}{\tilde \lambda_0 \tilde q_1} }&
1
\end{bmatrix}^T . 
$
Now~\eqref{eq:opfece_rfm} and~\eqref{eq:xsaxo} yield
$\tilde R=	
\frac{\tilde \lambda_0 \tilde\lambda_1  \tilde q_1}
{\tilde \lambda_0+
\tilde \lambda_1}  ,
$
	and
\begin{align*}
			\tilde e_1&=\frac{\tilde \zeta_3  }{\tilde\lambda_1^{1/2} \tilde  \kappa^* \tilde  \zeta_2} \\
			&=
			 \frac{\tilde \lambda_0 \tilde q_1 }{\tilde \lambda_0+\tilde \lambda_1} , 
\end{align*}
and this agrees with~\eqref{eq:ertyuo}
and~\eqref{eq:expeRR}.
\end{Example}

The spectral representation for the~{\rfmshort} can be applied to derive
  results on sensitivity analysis and quasi-concavity of the production rate. 

\begin{Corollary}\label{cor:rfmd_sens}
Consider an 
{\rfmshort} with dimension~$n$.
Let~$\tilde \kappa^*\geq 0$
denote  the unique solution of~$\sigma(\tilde A(\tilde \kappa))=\tilde \kappa$, and let~$\tilde \zeta^* \in \R^n_{++}$
denote the Perron  vector of~$\tilde A(\tilde \kappa^*)$
normalized such that~$(\tilde \zeta^*)^T\tilde \zeta^*=1$.
For any~$i\in\{1,\dots,n\}$ 
the  sensitivity of~$\tilde R $  with respect to a change of parameters  is given by
\begin{align}\label{eq:senselambda}
\frac{d}{d \tilde \lambda_{i-1} } \tilde R  &= 2 \left( 1- \sum_{i=1}^n (1-\tilde q_i)(\tilde \zeta^*_{i+1})^2  \right ) ^{-1}  \tilde \zeta^*_{i+1} \tilde \zeta^*_{i} \tilde \lambda_{i-1}^{-3/2} (\tilde k^*)^{-3},
\end{align}
and
\begin{align}\label{eq:diff_q}
\frac{d}{d \tilde q_i } \tilde R  &= 2  \left( 1- \sum_{i=1}^n (1-\tilde q_i)(\tilde \zeta^*_{i+1})^2  \right ) ^{-1}      (\tilde \zeta^*_{i+1})^2   (\tilde k^*)^{-2}. 
\end{align}
\end{Corollary}

\begin{Example}
Consider
the {\rfmshort} 
with $n=1$   and parameters~$\tilde \lambda_0=\tilde \lambda_1=1,\tilde q_1=0.5$.
Recall that in this case 
the Perron root of~$\tilde A(\tilde \kappa^*)$ is:
\[
\sigma (\tilde A(\tilde \kappa^*)) = \tilde \kappa^* =2 ,
\]
and the   steady-state flow rate is thus
\be
										R =	(\sigma (A( \kappa^*) ))^{-2} = 0.25.
\ee
The corresponding normalized Perron vector
is~$
				\tilde \zeta ^*=\frac{1}{\sqrt{6}}\begin{bmatrix}  1 & 2 & 1  \end{bmatrix}^T
$.
Calculating the sensitivity with respect to~$\tilde q_1$ using~\eqref{eq:senselambda}
yields
\begin{align}\label{eq:latpao_q1}
\frac{d}{d \tilde q_1 } \tilde R &= 2  \left( 1-   (1-\tilde q_1)(\tilde \zeta^*_{2})^2  \right ) ^{-1} (\tilde \zeta^*_{2})^2   (\tilde k^*)^{-2} \nonumber \\
&=   0.5. 
\end{align}

Let~$\varepsilon:=-0.01$ and suppose that 
 $\tilde q_1$ is decreased to~$ \tilde q_1+\varepsilon= 0.49 $. A direct calculation of the   optimal steady-state 
flow rate in the modified~{\rfmshort}  yields~$\overline  {\tilde R}=  0.245$,
so
\[
\frac{\overline  {\tilde R}-\tilde R}{\varepsilon}= \frac{-0.005}{-0.01} = 0.5 ,
\]
and this agrees well with~\eqref{eq:latpao_q1}.
 \end{Example}

\section{Conclusion}
The problem of modeling and analyzing the movement of ``biological machines'' 
along a 1D ``track''  is a central problem
in  systems biology.
Several  models have been proposed, both stochastic and deterministic. One recent
line of
research is related to the~RFM which is
 a deterministic model arising as an approximation to the more
fundamental stochastic model of~TASEP. The~TASEP describes an abstract assembly line, where the progress of the assembly
process is reflected by the forward motion of  particles along the linear sequence of assembly sites. Each particle attempts to hop
to the  the next site at random time, and if (and only if)
this next site is free the hop takes  place.

The~{\rfmshort} may be interpreted as 
   a mean-field dynamic approximation of  a   generalized~TASEP.
 Again, this is a model for an assembly line, where   the assembly
process is presented by a stochastic unidirectional motion of particles along a sequence of assembly sites. Again, each particle
tries to hop  forward to the next site at random time, but now this expected hop  is canceled   not only if the next
site is already occupied, but also if the next site 
is ``not ready'' 
to accept the particle. The ``readiness'' here is
described by  independent binary (ready/not ready) random variables 
 with probability~$q_i$ to be ready for site~$i$. 

Our results  show  that the dynamic 
mean-field approximation to this
 generalized~TASEP leads to a rich theory,  with
many powerful results.

A promising line of research is to study networks of interconnected~{\rfmshort}s
that can model the concurrent transport processes taking place in the cell. Another research direction  is the analysis of the corresponding
generalized~TASEP.  
Other  applications of the models introduced here are also
 of interest. 
For example, the~{\rfmshort} may be suitable for modeling vehicular traffic along a multi-lane road 
where the number of lanes changes along the road.


\section{Acknowledgments}
The work of MM is supported in part by research grants from  the~ISF 
and the~BSF. 
The work of AO was conducted in the framework of the state project
no.  AAAA-A17-117021310387-0 and is partially supported by RFBR grant
17-08-00742.
We are grateful to E. D. Sontag for helpful comments.


\appendix \label{sec:appendices1}

{\sl Proof of Prop.~\ref{prop:sta_inv}.}
The Jacobian matrix~$J(x)$ of~\eqref{eq:rfmrl} satisfies~$J(x)=M(x)-D(x)$ with
$D(x):=\diag \left (-\lambda_n  x_n - \lambda_1(q_2-x_2) , -\lambda_1  x_1 -\lambda_2 (q_3-x_3 ), \dots, 
-\lambda_{n-1}x_{n-1}  -\lambda_n  (q_1-x_1) \right )$ and $M(x)$ is
given in~\eqref{eq:long1}. 
\begin{figure*}
\begin{align}\label{eq:long1}
M(x):= \begin{bmatrix}
  0           &  \lambda_1 x_1  &0 &0&\dots &0&0  &\lambda_n(q_1-x_1) \\
\lambda_1 (q_2-x_2 )          &  0   &  \lambda_2 x_2 &0                                   & \dots&  0&0  & 0 \\
                          &              &                      & &\vdots\\
    0&0 &0&0&\dots&\lambda_{n-2}(q_{n-1}-x_{n-1}) &0& \lambda_{n-1}x_{n-1}\\
								 \lambda_n x_n& 0&0&0&\dots& 0&\lambda_{n-1}(q_n-x_n)   &0 
\end{bmatrix}.
\end{align}
\hrule
\end{figure*}
For any~$x\in C$ all the entries of~$M(x)$ are nonnegative, so the~{\rfmrshort} is a
cooperative   dynamical system~\cite{hlsmith}. 
Note that the matrix~$M(x)$ (and thus~$J(x)$)
 may become reducible 
for values~$x$ on the boundary of~$C$ e.g. for~$x$ such
 that~$x_2=q_2$ and~$x_n=0$. However,~$M(x)$ is irreducible 
for all~$x\in \Int(C)$.

Let~$0_n \in\R^n$ denote the vector with all entries     zero, 
and let~$q:=\begin{bmatrix}q_1 & q_2&\dots&q_n \end{bmatrix}^T$.
 Note that~$0_n$ and~$q $
are equilibrium points of the~{\rfmrshort}.
 For~$s=0$ and~$s=\sum_{i=1}^n q_i$
the corresponding level sets of~$H$ are~$L_0=\{0_n\}$ and~$L_{\sum_{i=1}^n q_i} =\{ q\}$  
and it is clear that for these values of~$s$ the proposition holds. 

Pick~$s\in(0,\sum_{i=1}^n q_i)$,
 and~$x(0)\in C$ such that~$\sum_{i=1}^n x_i(0)=s$. 
We claim that~$x(t)\in\Int(C)$ 
 for all~$t>0$. The proof of this follows from a cyclic version  
of~\cite[Lemma~1]{RFM_entrain} showing that~$C$ has 
 a repelling boundary.
The invariance result in Prop.~\ref{prop:sta_inv}
  follows from the fact that~$C$
is compact, convex and with a repelling boundary.

In particular, we conclude that 
for any~$t>0$ the matrix~$M(x(t))$
is irreducible, so the system is a cooperative irreducible system with~$H(x)$ as a first integral.
Now 
    the stability result in Prop.~\ref{prop:sta_inv}
		follows from the results in~\cite{mono_plus_int} (see also~\cite{Mierc1991}
    and~\cite{mono_chem_2007}
 for some related ideas).~$\square$

{\sl Proof of Prop.~\ref{prop:equi}.}
The proof is similar to the proof of~\cite[Prop.~1]{RFM_r_max_density}
and is therefore omitted.

{\sl Proof of Thm.~\ref{thm:pcfv}.}
Define~$f(\kappa)$ as in~\eqref{eq:deffk}.
Then~$f(0)=  \sigma( B)>0 $, and
\begin{align*}
											&\lim_{\kappa\to\infty} f(\kappa) \\			
											&=\lim_{\kappa\to\infty} \sigma\Big(\kappa\diag ( 1-q_1,1-q_2,...,1-q_n )  \Big)-\kappa\\
											&=\lim_{\kappa\to\infty}   \Bigg(  \sigma\Big(\diag ( 1-q_1,1-q_2,...,1-q_n )  \Big)-1 \Bigg)\kappa\\
											&=-\infty,
\end{align*} 
because~$1-q_i<1$ for all~$i$. 
By continuity, we conclude that there  
  exists a value~$ \kappa^*>0$ such that~$ f( \kappa^*)=0$, i. e.
\[
\sigma\Big(\kappa^*\diag ( 1-q_1,1-q_2,...,1-q_n ) +B\Big)= \kappa^*.
\]

We now show that the value~$ \kappa^*$ is unique. For~$\kappa \in [0,\infty)$, let~$\zeta(\kappa)   $ denote the
normalized  Perron  vector of the componentwise nonnegative and irreducible matrix~$A(\kappa  )  $,
i.e.~$ \zeta(\kappa) \in\R^n_{++} $ and~$\zeta^T(\kappa)\zeta(\kappa)=1$. 
Then using known results for the sensitivity of the Perron eigenvalue (see, e.g.~\cite{magnus85}) and the fact that~$A(\kappa) $
is symmetric
 yields 
\begin{align*}
			\frac{d}{d\kappa}f(\kappa)&=  \zeta^T(k) \diag ( 1-q_1,1-q_2,...,1-q_n )    \zeta(k) -1\\
			&\leq \max_i \{1-q_i\}\zeta^T(k) \zeta(k)-1\\
			&\leq -\ell, 
\end{align*}
where~$\ell:=\min_i {q_i}>0$. Note that~$\ell$ does not depend on the rates.
Thus,~$f(\kappa)$ is strictly  decreasing in~$\kappa$, implying that~$ \kappa^*$ is unique.

To prove the spectral representation, consider
the~$n\times n$ periodic Jacobi matrix
\begin{align*}
F:=\begin{bmatrix}
p_1 & c_1 & 0 & 0 & \dots & 0 & c_n \\
c_1 & p_2 & c_2 & 0 & \dots & 0 & 0 \\
0 & c_2 & p_3 & c_3 & \dots & 0 & 0 \\
&&&\vdots \\
0 & 0 & 0 & \dots & c_{n-2} & p_{n-1} & c_{n-1}  \\
c_n & 0 & 0 & \dots & 0 & c_{n-1} & p_n \end{bmatrix},
\end{align*}
with~$c_i>0$ and~$p_i\geq 0$ for all~$i$. 
Since~$F$ is componentwise nonnegative and irreducible, it admits a Perron root~$\sigma>0$
and a Perron vector~$\zeta\in\R^n_{++}$. 
The equation~$F\zeta=\sigma\zeta$ gives
\begin{align}\label{eq:psdco} 
p_i \zeta_i + c_i \zeta_{i+1} + c_{i-1} \zeta_{i-1} = \sigma \zeta_i,\quad i=1,\dots,n,  
\end{align}
where all the indexes here and below are modulo~$n$. 
Let
\begin{align}\label{eq:defdi}
d_i := \frac{c_i\zeta_{i+1}}{\sigma \zeta_{i}} ,\quad i=1,\dots,n.
\end{align}
Note that~$d_i>0$ for all~$i$, and 
that
\be\label{eq:prodeq}
 \prod_{i=1}^{n}d_i   
= \sigma^{-n} \prod_{i=1}^{n}c_i.
\ee

Eq.~\eqref{eq:psdco} yields 
\begin{align*}
\frac{c_{i-1} \zeta_{i-1}}{\sigma \zeta_i} =  1-\frac{p_i}{\sigma} -d_i,\quad i=1,\dots,n.
\end{align*}
Multiplying both sides of this equation 
 by~$d_{i-1}$ and rearranging gives:
\begin{align}\label{eq:alkgyrtr}
 \sigma^{-2} &= c_{i-1}^{-2}\left(    1-\frac{p_i}{\sigma}   -d_i \right ) d_{i-1},
\quad i=1,\dots,n.
\end{align} 
This implies that
\[
\sigma^{-2n}\prod_{i=1}^n c_i^2 =  \prod_{i=1}^n 
d_i  \prod_{i=1}^n    \left(    1-\frac{p_i}{\sigma}   -d_i \right ), 
\]
and combining this with~\eqref{eq:prodeq} gives
\be\label{eq:diprod}
 \prod_{i=1}^n d_i  =    \prod_{i=1}^n    \left(    1-\frac{p_i}{\sigma}   -d_i \right ). 
\ee

To relate this to the~{\rfmrshort}, 
note that the matrix~$A(\kappa^*)$
 has the same form as~$F$ with
\[
			p_i= (1-q_i)\kappa^*,\;
			c_i=\lambda_i^{-1/2},\; \sigma=\kappa ^*,
\]
and then
$
			1-\frac{p_i}{\sigma}=q_i , 
$
so~\eqref{eq:alkgyrtr} and~\eqref{eq:diprod} 
become
\begin{align*}
(k^*)^{-2} &=\lambda_{i-1}  d_{i-1}
\left(    q_i   -d_i \right ),
\quad i=1,\dots,n,\\
 \prod_{i=1}^n d_i & =    \prod_{i=1}^n    \left(  q_i  -d_i \right ).
\end{align*}
Comparing this with~\eqref{eq:ewqqp} and~\eqref{eq:lgtpp}, that admit  a unique solution, 
we conclude that~$R^*=(k^*)^{-2}$ and~$ e_i^*=d_i$
for all~$i$.
Applying~\eqref{eq:defdi} completes the proof of
Thm.~\ref{thm:pcfv}.~$\square$

{\sl Proof of Corollary~\ref{coro:vobu}.}
Pick~$m \in \{1,\dots,n\}$. Consider the~{\rfmrshort}
with~$\lambda_m\to \infty$. Thus,~$B(\lambda)\to \bar B$, where in~$\bar B$
 the  entries~$\lambda_m^{-1/2}$ are replaced by zero.
It is straightforward to see that~$\bar B$ is 
  componentwise nonnegative and irreducible.  It follows from the proof 
of Thm.~\ref{thm:pcfv} that a unique~$\kappa^* \in [0,\infty)$
exists and then by continuity we conclude that~$R^*$ and~$e^*$ can be obtained from the
Perron root, which is simple,
 and the corresponding Perron vector
 of~$\bar B$.~\hfill{$\square$}


{\sl Proof of Prop.~\ref{prop:Rsens_lambda_and_q}.} 
Recall that~$A(\kappa ):=\kappa D(q)+B(\lambda)$
 (see~\eqref{eq:pjm}), 
and~$f(\kappa,D,B):=\sigma(A(\kappa))-\kappa$. 
The value~$\kappa^*(D,B)$ is the unique value
such that~$f(\kappa^*(D,B),D,B)=0$,
and~$(R^*)^{-1/2}=\kappa^*$. 
To simplify the notation, we write~$C$ for the pair of matrices~$(D,B)$.
Suppose that~$p$ is a parameter in~$C$.
Our goal is to determine the sensitivity 
\[
\frac{d}{dp} f(\kappa^*(C(p)),C(p)) .
\]

Differentiating the equation~$f(\kappa^*(C(p)),C(p))=0$
with respect to~$C$ yields
\begin{align*}
0&=\left( \frac{d}{d \kappa } f \right ) \left( \frac{d}{dC } \kappa^*\right )  +\left( \frac{d}{dC } f \right ) \\
&=\left( \frac{d}{d \kappa } \sigma(\kappa^*) -1 \right ) \left( \frac{d}{dC } \kappa^*\right )  +\frac{d}{dC } f.
\end{align*}

We know
 that~$  \frac{d}{d \kappa } \sigma(\kappa^*) -1 <0$, so in particular it is not zero and
\be\label{eq:tnbuysla}
\frac{d}{dC }  \kappa^*  =  
\left( 1-\frac{d}{d\kappa } \sigma(\kappa^* )   \right ) ^{-1}   \frac{d}{dC } \sigma(\kappa^*)   .
\ee

We now consider two cases.

\noindent {\sl Case 1.}  Suppose that~$p=\lambda_i$ for some~$i$. Recall that~$D=D(q)$ and~$B=B(\lambda )$, so
\begin{align*}
\frac{d}{d \lambda_i }\kappa^*&=  \left( \frac{d}{d C } \kappa^*   \right ) \left(\frac{d}{d \lambda_i } C  \right ) \\
& =  \left( 1-\frac{d}{d \kappa } \sigma(\kappa^* )   \right ) ^{-1}   \frac{d}{dC } \sigma(\kappa^* ) \frac{d}{d \lambda_i } C \\
& =  \left( 1-\frac{d}{d\kappa } \sigma(\kappa^*D+B)   \right ) ^{-1}   \frac{d}{d \lambda_i  } \sigma(\kappa^*D+B)  ,
\end{align*}
where the second equation follows from~\eqref{eq:tnbuysla}.
Let~$\zeta^* $ denote the Perron  vector of the symmetric matrix~$k^*D+B$, normalized
so that~$(\zeta^*)^T \zeta^*=1$.  Then
\begin{align*}
\frac{d}{d \lambda_i } \kappa^*&= \left( 1-(\zeta^*)^T D   \zeta^*  \right ) ^{-1} (\zeta^*)^T \frac{d B}{d\lambda_i } \zeta^* \\
&=  \left( 1- \sum_{i=1}^n (1-q_i)(\zeta^*_i)^2  \right ) ^{-1} ( -\zeta_i^* \zeta_{i+1}^* \lambda_i^{-3/2} ) , 
\end{align*}
where the last equation follows from the definitions of~$D$ and~$B$. Using the fact that~$R^*=(\kappa^*)^{-2}$
yields~\eqref{eq:senselam}.
 
\noindent {\sl Case 2.} 
Suppose that~$p=q_i$ for some~$i$. Then
\begin{align*}
\frac{d}{d q_i } \kappa^*&=  \left( \frac{d}{d C } \kappa^*   \right ) \left(\frac{d}{d q_i } C  \right ) \\
& =  \left( 1-\frac{d}{d\kappa } \sigma(\kappa^* )   \right ) ^{-1}   \frac{d}{dC } \sigma(\kappa^* ) \frac{d}{d q_i } C \\
& =  \left( 1-\frac{d}{d\kappa } \sigma(\kappa^*D+B)   \right ) ^{-1}   \frac{d}{d q_i  } \sigma(\kappa^*D+B)  .
\end{align*}
Thus,
\begin{align*}
\frac{d}{d q_i } \kappa^*&= \left( 1-(\zeta^*)^T D   \zeta^*  \right ) ^{-1} (\zeta^*)^T \frac{d (\kappa^* D )}{d q_i} \zeta^* \\
&= -\kappa^* \left( 1- \sum_{i=1}^n (1-q_i)(\zeta^*_i)^2  \right ) ^{-1} (\zeta_i^*)^2,
\end{align*}
and combining this 
with the fact that~$R^*=(\kappa^*)^{-2}$
yields~\eqref{eq:dfvcf}.
This completes the proof of
  Prop.~\ref{prop:Rsens_lambda_and_q}.~$\square$

 {\sl Proof of Thm.~\ref{thm:R^(-1/2)_convex}.}
For a symmetric matrix~$S \in \R^{n\times n}$,
let~$\lambda_{\max}(S)\in\R$ denote
 the maximal eigenvalue of~$S$. 
Recall that the~$L_2$ induced matrix norm is~$||A||_2=(\lambda_{\max} (A^T A ) )^{1/2}$. If~$A$ is symmetric and componentwise nonnegative then this gives
\begin{align*}
||A||_2&=(\lambda_{\max} (A^2 ) )^{1/2}\\
&=\sigma(A),
\end{align*}
where~$\sigma(A)$ is the Perron root of~$A$. 
Since any matrix norm is convex, 
this implies that  the Perron root is convex over the 
set of 
symmetric and componentwise nonnegative matrices. 

Pick~$r\in[0,1]$,
$p=\begin{bmatrix}  \lambda&q \end{bmatrix}^T$,
$\tilde p=\begin{bmatrix}\tilde \lambda& \tilde q \end{bmatrix}^T$, such that~$p,\tilde p\in \left(\R^n_{++} \times (0,1]^n\right)$, and let
$\bar B:=r B(\lambda)+(1-r) B(\tilde \lambda)$ and
$\bar D:=r D(q)+(1-r) D(\tilde q)$.
 Then for any~$\kappa\geq 0$ we have 
\begin{align}\label{eq:lpcod}
&f(\kappa,  \bar D , \bar B )= \sigma( \kappa \bar D+ \bar B) -\kappa \nonumber \\
&\leq r\sigma(\kappa D(q)+B(\lambda))+(1-r) \sigma( k   D(\tilde q)+ B(\tilde \lambda) )-k \nonumber \\
&=r f(\kappa,   D(q) ,   B(\lambda) )+(1-r) 
 f(\kappa,   D(\tilde q) ,   B(\tilde \lambda) ),
\end{align}
where the second equation follows from the convexity of~$\sigma$. 

 Seeking a contradiction, suppose that
\[
\kappa^*(  \bar D,\bar B ) >  \max \{  \kappa^*(D(q),B(\lambda) ), \kappa^*(D(\tilde q),B(\tilde \lambda)) \}.
\]
Since~$f$ decreases with~$\kappa$, this yields
\begin{align*}
f( \kappa^*(  \bar D,\bar B ), D(q),B(\lambda) ) &<0 \\
 f(\kappa^*(  \bar D,\bar B ), D(\tilde q),B(\tilde \lambda) )&<0, 
\end{align*}
and combining  this with~\eqref{eq:lpcod} gives
\[
f( \kappa^*(  \bar D,\bar B ),  \bar D , \bar B ) <0.
\]
However, this contradicts the definition of~$\kappa^*(  \bar D,\bar B )$. We conclude that
\[
\kappa^*(  \bar D,\bar B )\leq \max \{  \kappa^*(D(q),B(\lambda) ), \kappa^*(D(\tilde q),B(\tilde \lambda)) \},
\]
and using the fact that~$R^*=(\kappa^*)^{-2}$ gives 
\[
R^*(  \bar D,\bar B )\geq \min \{ R^*(D(q),B(\lambda) ), R^*(D(\tilde q),B(\tilde \lambda)) \}.
\]
This proves~\eqref{eq:rcomnac}.

To complete the proof, 
let~$\kappa^*(q,\lambda):=\kappa^*(D(q),B(\lambda))$,
 so that~$\kappa^* (q,\lambda) = \sigma(\kappa^*  (q,\lambda) D(q)+B(\lambda)) $.
 Fix~$c>0$. 
Then  clearly 
$
					c^{-1/2} \kappa^*  (q,\lambda) = \sigma(c^{-1/2} \kappa^*  (q,\lambda) D(q)+c^{-1/2}B(\lambda)) 
					$,
and using the definition of~$B(\lambda)$ yields
\[
					c^{-1/2} \kappa^*  (q,\lambda)  = \sigma(c^{-1/2} \kappa^*  (q,\lambda) D(q)+ B(c \lambda)) .
\]

We conclude that
\[
     \kappa^*(  D(q), B(c \lambda) )=c^{-1/2} \kappa^*(D(q),B(\lambda)) ,
\]
so
\[
			R^*(q,c\lambda)=c R^*(q,\lambda).
\]
In other words, for a fixed~$q$ the optimal steady-state flow rate
is homogeneous of degree one with respect to~$\lambda$. 
Combining this with~\eqref{eq:rcomnac}
 completes the proof of 
Thm.~\ref{thm:R^(-1/2)_convex}.~$\square$

{\sl Proof of Thm.~\ref{thm:period}.}
Let~$0_n \in\R^n$ denote the vector with all entries     zero, 
and let~$q:=\begin{bmatrix}q_1 & q_2&\dots&q_n \end{bmatrix}^T$.
 Note that~$0_n$ and~$q $
are equilibrium points of the~{\prfmrshort} (and thus they
 are~$T$-periodic solutions).
 For~$s=0$ and~$s=\sum_{i=1}^n q_i$
the corresponding level sets of~$H$ are~$L_0=\{0_n\}$ and~$L_{\sum_{i=1}^n q_i} =\{ q\}$  
and it is clear that for these values of~$s$ the theorem holds. 

Pick~$s\in(0,\sum_{i=1}^n q_i)$,
 and~$x(0)\in C$ such that~$\sum_{i=1}^n x_i(0)=s$. 
We know from the proof of Prop.~\ref{prop:sta_inv} that~$x(t)\in\Int(C)$ 
 for all~$t>0$. Now the entrainment result follows 
from~\cite[Theorem~A]{JI-FA96}.~\hfill{$\square$}

{\sl Proof of Prop.~\ref{prop:rfmd_sta_inv}.}
The proof is similar to the proof of Prop.~\ref{prop:sta_inv}
and is therefore omitted.

{\sl Proof of Prop.~\ref{prop:rfmrd_to_rfmd}.}
Recall that the optimal-density steady-state in an~$(n+2)$-dimensional~{\rfmrshort} 
satisfies
\be\label{eq:ipor}
\prod_{i=1}^{n+2} e^*_i = \prod_{i=1}^{n+2} (q_i-e^*_i),
\ee
and
\begin{align}\label{eq:poydd} 
 \lambda_{n+2} e_{n+2}^*(q_1-e_1^*)  &=\lambda_1 e_1^*(q_2-e_2^*) \nonumber \\
    &=\lambda_2 e^*_2(q_3-e^*_3) \nonumber \\
		&\vdots \nonumber \\
		&=\lambda_{n} e^*_{n}(q_{n+1}-e^*_{n+1})  \nonumber \\
		&=\lambda_{n+1} e^*_{n+1}(q_{n+2}-e^*_{n+2})  .
\end{align}
Suppose that~$\lambda_{n+2} \to \infty$. We know from
 Corollary~\ref{coro:vobu} that the~$e^*_i$'s
 remain bounded, so
\eqref{eq:poydd}  implies that~$e_{n+2}^*(q_1-e_1^*) \to 0 $.
This implies that at least one of the two terms~$e_{n+2}^*  $,
$q_1-e_1^*$ goes to zero.  We consider these two cases. We will show that in both cases \emph{both}~$e_{n+2}^*$
and~$q_1-e_1^*$ go to zero.

\noindent {\sl Case 1.} Suppose that~$e_{n+2}^* \to 0$.
 Then~\eqref{eq:ipor} implies that there exists~$i\in\{1,\dots,n+1\}$
such that~$e^*_i \to q_i$.  Seeking a contradiction, assume that
\begin{align}\label{eq:cont_case1}
e^*_1\not\to  q_1
\end{align}
Then there exists~$i\in\{2,\dots,n+1\}$
such that~$e^*_i \to q_i$.  
 Now~\eqref{eq:poydd}  gives~$e_{n+1}^*  \to 0$. 
Applying~\eqref{eq:poydd}  again
gives~$e_{n}^*  \to 0$, and proceeding in this way gives~$e_i^*\to 0$ for~$i=1,\dots,n+1$.  Substituting this in~\eqref{eq:ipor}
yields
$0 =   q_1   \dots  q_{n+2}
$, and this is impossible as we assume that~$q_i>0$ for all~$i$.

\noindent {\sl Case 2.} Suppose that~$e^*_1\to q_1$.
Eq.~\eqref{eq:ipor} gives 
$\left(\prod_{i=1}^{n+2} e^*_i\right) \to 0$. Seeking a contradiction,
assume that
\begin{align}\label{eq:cont_case2}
e^*_{n+2} \not \to 0
\end{align}
Then
 there exists an index~$i\in\{1,\dots,n+1\}$
such that~$e^*_i \to 0$. 
Now~\eqref{eq:poydd} implies that~$e^*_2\to q_2$.
Using~\eqref{eq:poydd} again gives~$e^*_3\to q_3$, 
and proceeding in this fashion we conclude that~$e^*_i \to q_i$ for~$i=1,\dots, n+2$. But this contradicts~\eqref{eq:cont_case2}. We conclude that~$e^*_{n+2} \to 0$.

Summarizing, we showed that when~$\lambda_{n+2}\to\infty$
both~$e_1^*\to q_1$ and~$e_{n+2}^*\to 0$. 
Substituting this in~\eqref{eq:poydd} gives
\begin{align}\label{eq:agfter}
  \lambda_1 q_1 (q_2-e_2^*) 
    &=\lambda_2 e^*_2(q_3-e^*_3) \nonumber \\
		&\vdots \nonumber \\
		&=\lambda_{n} e^*_{n}(q_{n+1}-e^*_{n+1})  \nonumber \\
		&=\lambda_{n+1} e^*_{n+1}q_{n+2}   .
\end{align}

Consider the steady-state equations for an~$n$-dimensional~{\rfmshort}, that is, 
\begin{align*}\label{eq:eqrfmrd}
    \tilde \lambda_0 ( \tilde q_1- \tilde e_1) &=  \tilde\lambda_1  \tilde  e_1( \tilde q_2- \tilde  e_2) \\
		&=  \tilde\lambda_{2}  \tilde  e_{2} ( \tilde q_3- \tilde e_3)   \nonumber \\
                             &\vdots \nonumber \\
                   &= \tilde\lambda_{n-1} \tilde e_{n-1} ( \tilde q_n- \tilde  e_n)\\
									&= \tilde \lambda_n  \tilde  e_n. 
\end{align*} 
Comparing this to~\eqref{eq:agfter}, we conclude that 
if~\eqref{eq:addrates} and~\eqref{eq:addqia}
hold then the two sets of
equations are identical up to the replacement~$\tilde e_i=e_{i+1}^*$ for all~$i$. Since both sets of equations admit a unique feasible solution, this proves~\eqref{eq:propr77}.~\hfill{$\square$}


{\sl Proof of Corollary~\ref{cor:rfmd_spectral}.}
We   construct    a corresponding~$(n+2)$-dimensional {\rfmrshort}
as described in 
 Remark~\ref{rem:copos}. Recall that this has rates
and site sizes given  in~\eqref{eq:rem2rates} and~\eqref{eq:rem2sir},
with~$a>0$, and that we use~$e(a)$
to denote the steady-state of this~{\rfmrshort}.
By Thm.~\ref{thm:pcfv},
the spectral representation for the optimal-density steady-state of this~{\rfmrshort}
is based on the~$(n+2)\times(n+2)$ matrix
\be 
A(\kappa) =\left [\begin{smallmatrix}
0 & \tilde  \lambda_0^{-1/2}   & 0 &0 & \dots &0&a^{-1/2} \\
\tilde \lambda_0^{-1/2} & (1-\tilde q_1)\kappa & \tilde \lambda_1^{-1/2}   & 0  & \dots &0&0 \\
 0&\tilde \lambda_1^{-1/2} & (1-\tilde q_2)\kappa &  \tilde \lambda_2^{-1/2}    & \dots &0&0 \\
 & &&\vdots \\
 0& 0 & 0 & \dots &\tilde \lambda_{n-2}^{-1/2}  & (1-\tilde  q_{n})\kappa & \tilde \lambda_{n}^{-1/2}     \\
a^{-1/2}& 0 & 0 & \dots & 0 & \tilde\lambda_{n }^{-1/2}  & 0
 \end{smallmatrix} \right ],
\ee
where we used the fact that~$q_1=q_{n+2}=1$.
Note that~$A(\kappa)$ is componentwise nonnegative and irreducible
 for all~$\kappa\geq 0$. 
By Thm.~\ref{thm:pcfv},
there exists a unique value~$ \kappa^*\in [0,\infty)$ such that the matrix~$A(\kappa)$
satisfies
\[\sigma(A( \kappa^*))=\kappa^* . 
\]
and
 the optimal steady-state densities and flow rate satisfy
\begin{align}\label{eq:cfovv}
					e^*_i&=\frac{\zeta_{i+1}(  \kappa^*)}{\lambda_i^{1/2} \sigma (  \kappa^*)  \zeta_{i}( \kappa^*)} \nonumber \\
					&=\frac{\zeta_{i+1}(  \kappa^*)}{\tilde \lambda_{i-1}^{1/2} \sigma (  \kappa^*)  \zeta_{i}( \kappa^*)}  
\end{align}
for every~$i$, 
and $R^*=	(\sigma (  \kappa^*) )^{-2}$.

When~$a\to \infty$,~$A(\kappa)$ converges to the matrix~$\tilde A(\kappa)$ in~\eqref{eq:atildd}. 
 Eq.~\eqref{eq:ealimi}
implies that
for any~$i=1,\dots,n$ we have  
\[
\tilde e_i =\lim_{a\to\infty} e^*_{i+1}(a).
\]
Combining this with~\eqref{eq:cfovv} and continuity  of the Perron root (which is a simple eigenvalue) and Perron vector
completes the proof.~\hfill{$\square$}


{\sl Proof of Corollary~\ref{cor:rfmd_sens}.} 
Recall that~$\tilde A(\tilde \kappa ):=\tilde \kappa \tilde D(\tilde q)+\tilde B(\tilde \lambda)$, where~$\tilde D(\tilde q)$ is the diagonal matrix with entries
$ 0,1-\tilde q_1,1-\tilde q_2,...,1-\tilde q_n,0$ on the diagonal, 
and~$\tilde f(\tilde \kappa,\tilde D,\tilde B):=\sigma(\tilde A(\tilde \kappa))-\tilde \kappa$. 
The value~$\tilde \kappa^*(\tilde D,\tilde B)$ is the unique value
such that~$\tilde f(\tilde \kappa^*(\tilde D,\tilde B),\tilde D,\tilde B)=0$,
and~$(\tilde R^*)^{-1/2}=\tilde \kappa^*$. 
To simplify the notation, we write~$\tilde C$ for the pair of matrices~$(\tilde D,\tilde B)$.
Suppose that~$\tilde p$ is a parameter in~$\tilde C$.
Our goal is to determine the sensitivity 
\[
\frac{d}{d\tilde p} \tilde f(\tilde \kappa^*(\tilde C(\tilde p)),\tilde C(\tilde p)) .
\]

Differentiating the equation~$\tilde f(\tilde \kappa^*(\tilde C(\tilde p)),\tilde C(\tilde p))=0$
with respect to~$\tilde C$ yields
\begin{align*}
0&=\left( \frac{d}{d \tilde \kappa } \tilde f \right ) \left( \frac{d}{d \tilde C } \tilde \kappa^*\right )  +\left( \frac{d}{d \tilde C } \tilde f \right ) \\
&=\left( \frac{d}{d \tilde \kappa } \sigma(\tilde \kappa^*) -1 \right ) \left( \frac{d}{d \tilde C } \tilde \kappa^*\right )  +\frac{d}{d \tilde C } \tilde f.
\end{align*}

We know
 that~$  \frac{d}{d \tilde \kappa } \sigma(\tilde \kappa^*) -1 <0$, so in particular it is not zero and
\be\label{eq:tnbuyslatild}
\frac{d}{d \tilde C }  \tilde \kappa^*  =  
\left( 1-\frac{d}{d \tilde \kappa } \sigma(\tilde \kappa^* )   \right ) ^{-1}   \frac{d}{d \tilde C } \sigma(\tilde \kappa^*)   .
\ee

We now consider two cases.

\noindent {\sl Case 1.}  Suppose that~$\tilde p=\tilde \lambda_{i-1}$ for~$i\in\{1,2,\hdots,n+1\}$. Recall that~$\tilde D=\tilde D(\tilde q)$ and~$\tilde B=\tilde B(\tilde \lambda )$, so
\begin{align*}
&\frac{d}{d \tilde \lambda_{i-1} }\tilde \kappa^*=  \left( \frac{d}{d \tilde C } \tilde \kappa^*   \right ) \left(\frac{d}{d \tilde \lambda_{i-1} } \tilde C  \right ) \\
& =  \left( 1-\frac{d}{d \tilde \kappa } \sigma(\tilde \kappa^* )   \right ) ^{-1}   \frac{d}{d \tilde C } \sigma(\tilde \kappa^* ) \frac{d}{d \tilde \lambda_{i-1} } \tilde C \\
& =  \left( 1-\frac{d}{d \tilde \kappa } \sigma(\tilde \kappa^* \tilde D+\tilde B)   \right ) ^{-1}   \frac{d}{d \tilde \lambda_{i-1}  } \sigma(\tilde \kappa^* \tilde D+\tilde B)  ,
\end{align*}
where the second equation follows from~\eqref{eq:tnbuyslatild}.
Let~$\tilde \zeta^* $ denote the Perron  vector of the symmetric matrix~$\tilde k^* \tilde D+\tilde B$, normalized
so that~$(\tilde \zeta^*)^T \tilde \zeta^*=1$.  Then
\begin{align*}
&\frac{d}{d \tilde \lambda_{i-1} } \tilde \kappa^*= \left( 1-(\tilde \zeta^*)^T \tilde D \tilde \zeta^* \right ) ^{-1} (\tilde \zeta^*)^T \frac{d \tilde B}{d\tilde \lambda_{i-1} } \tilde \zeta^* \\
&=  \left( 1- \sum_{i=1}^n (1-\tilde q_i)(\tilde \zeta^*_{i+1})^2  \right ) ^{-1} ( -\tilde \zeta_{i+1}^* \tilde \zeta_{i}^* \tilde \lambda_{i-1}^{-3/2} ) , 
\end{align*}
where the last equation follows from the definitions of~$\tilde D$ and~$\tilde B$. Using the fact that~$\tilde R^*=(\tilde \kappa^*)^{-2}$
yields~\eqref{eq:senselambda}.
 
\noindent {\sl Case 2.} 
Suppose that~$\tilde p=\tilde q_i$ for some~$i$. Then
\begin{align*}
&\frac{d}{d \tilde q_i } \tilde \kappa^*=  \left( \frac{d}{d \tilde C } \tilde \kappa^* \right ) \left(\frac{d}{d \tilde q_i } \tilde C  \right ) \\
& =  \left( 1-\frac{d}{d \tilde \kappa } \sigma(\tilde \kappa^* )   \right ) ^{-1}   \frac{d}{d\tilde C } \sigma(\tilde \kappa^* ) \frac{d}{d \tilde q_i } \tilde C \\
& =  \left( 1-\frac{d}{d\tilde \kappa } \sigma(\tilde \kappa^* \tilde D+\tilde B)   \right ) ^{-1}   \frac{d}{d \tilde q_i  } \tilde \sigma(\tilde \kappa^* \tilde D+\tilde B)  .
\end{align*}
Thus,
\begin{align*}
\frac{d}{d \tilde q_i } \tilde \kappa^*&= \left( 1-(\tilde \zeta^*)^T \tilde D  \tilde \zeta^*  \right ) ^{-1} (\tilde \zeta^*)^T \frac{d (\tilde \kappa^* \tilde D )}{d \tilde q_i} \tilde \zeta^* \\
&= -\tilde \kappa^* \left( 1- \sum_{i=1}^n (1-\tilde q_i)(\tilde \zeta^*_{i+1})^2  \right ) ^{-1} (\tilde \zeta_{i+1}^*)^2,
\end{align*}
and combining this 
with the fact that~$\tilde R^*=(\tilde \kappa^*)^{-2}$
yields~\eqref{eq:diff_q}.
This completes the proof of
  Corollary~\ref{cor:rfmd_sens}.~$\square$






\end{document}